\documentclass[]{jfm}
\usepackage[latin1]{inputenc}
\usepackage[english]{babel}
\usepackage{graphicx}
\usepackage{subfig}
\usepackage{array}
\usepackage{amsmath}
\usepackage{SIunits}
\usepackage{natbib}
\usepackage{color}
\graphicspath{{figspdflatex/}}

\newcommand{\pd}[2]{\frac{\partial #1}{\partial #2}}
\newcommand{\nd}[2]{\frac{d #1}{d #2}}

\title{Collapse and pinch-off of a non-axisymmetric impact-created air cavity in water}
\author[O. Enriquez, I. Peters, S. Gekle, L. Schmidt, D. Lohse, and D. van der Meer]{O\ls S\ls C\ls A\ls R\ns R.\ns E\ls N\ls R\ls I\ls Q\ls U\ls E\ls Z,\ns I\ls V\ls O\ns R.\ns P\ls E\ls T\ls E\ls R\ls S, S\ls T\ls E\ls P\ls H\ls A\ls N\ns G\ls E\ls K\ls L\ls E,\ns L\ls A\ls U\ls R\ls A\ns E.\ns S\ls C\ls H\ls M\ls I\ls D\ls T, \ns D\ls E\ls T\ls L\ls E\ls F\ns L\ls O\ls H\ls S\ls E, \and D\ls E\ls V\ls A\ls R\ls A\ls J\ns V\ls A\ls N\ns D\ls E\ls R\ns M\ls E\ls E\ls R\ns}
\affiliation{University of Twente, Enschede, The Netherlands}
\date{\today}

\begin{document}
\maketitle


\begin{abstract}
The axisymmetric collapse of a cylindrical air cavity in water follows a universal power law with logarithmic corrections. Nonetheless, it has been suggested that the introduction of a small azimuthal disturbance induces a long term memory effect, reflecting in oscillations which are no longer universal but remember the initial condition. In this work, we create non-axisymmetric air cavities by driving a metal disc through an initially-quiescent water surface and observe their subsequent gravity-induced collapse. The cavities are characterized by azimuthal harmonic disturbances with a single mode number $m$ and amplitude $a_m$. For small initial distortion amplitude (1 or 2 \% of the mean disc radius), the cavity walls oscillate linearly during collapse, with nearly constant amplitude and increasing frequency. As the amplitude is increased, higher harmonics are triggered in the oscillations and we observe more complex pinch-off modes.  For small amplitude disturbances we compare our experimental results with the model for the amplitude of the oscillations by \citet{Schmidt_Nat_09} and the model for the collapse of an axisymmetric impact-created cavity previously proposed by \citet{Bergmann_JFM_09}. By combining these two models we can reconstruct the three-dimensional shape of the cavity at any time before pinch-off. 
\end{abstract}


\section{Introduction}
\label{intro}

The pinch-off of an axisymmetric air cavity in water  is characterized by a finite-time singularity. The kinetic energy of the flow is focused into a vanishing small volume with a velocity whose magnitude diverges as the pinch-off moment is approached. Several experimental and theoretical scenarios have been recently considered in the study of this problem:  a bubble rising from a capillary \citep*{Longuet_JFM_91, Oguz_JFM_93, Burton_PRL_05, Thoroddsen_POF_07}, bubbles in a co-flowing liquid \citep*{Gordillo_PRL_05, Bergmann_PRL_09}, an initially necked bubble \citep*{Eggers_PRL_07}, and cavities created through impact \citep*{Bergmann_PRL_06, Gekle_PRL_08, Bergmann_JFM_09}. Depending on the case, the collapse might be initiated by surface tension, external flow, or hydrostatic pressure. However, irrespective of the cause, towards the end it is the inertia of the fluid that takes over in every case, and the collapse is accelerated as the radius of the cavity shrinks.

The time it takes each of these systems to reach the inertial collapse regime varies by orders of magnitude \citep{Gekle_PRE_09}. Hence, it was not an easy task to determine whether there was indeed a universal behaviour underlying this phenomenon. The first proposed model was a power law where the radius decreased proportionally to the square root of the remaining time until collapse, $\tau$ \citep{Longuet_JFM_91, Oguz_JFM_93}. Subsequent experimental and numerical studies consistently found the behaviour deviated slightly from that $\tfrac{1}{2}$ power law \citep{Burton_PRL_05, Gordillo_PRL_05, Keim_PRL_06, Bergmann_PRL_06, Thoroddsen_POF_07, Bergmann_JFM_09}, generating doubts and starting a controversy about the universality of the phenomenon.   \citet{Gordillo_JFM_06} and \citet{Eggers_PRL_07} theoretically showed how the power law varies weakly as a function of $\tau$ due to a logarithmic correction.

In conclusion, the axisymmetric problem converges to a universal self-similar solution. If axial symmetry is broken by a small azimuthal perturbation, a truly universal system would be expected to converge to the same solution . However, through experiments and simulations of an air bubble disconnecting from an underwater nozzle,  \citet*{Keim_PRL_06, Schmidt_Nat_09,Turitsyn_PRL_09}; and \citet*{Keim_PRE_11}  recently showed that a slight azimuthal asymmetry can trigger vibrations that persist in time.  The fact that a small perturbation is not smoothed out indicates that the system possesses memory of its initial conditions. Here we conduct an experimental study of the evolution of azimuthal disturbances in the collapse of an otherwise axisymmetric cavity.

\subsection*{Impact vs. detachment} 
\begin{figure}
\centering
\includegraphics[]{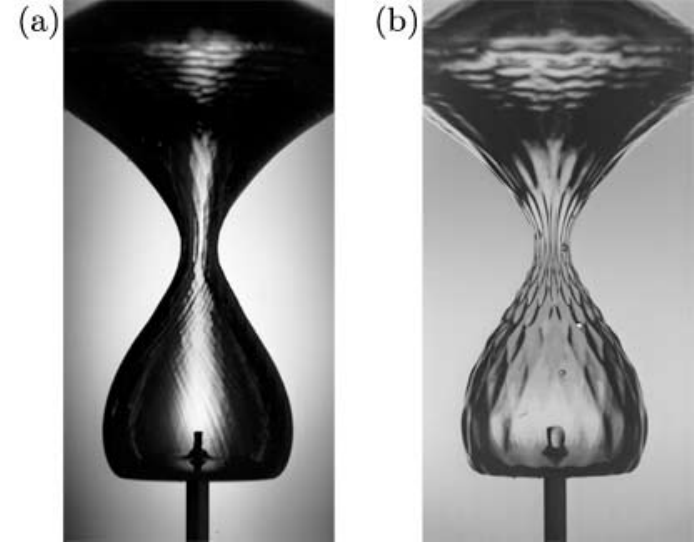}
\caption{Collapse of a cavity created through impact of a round disc (a) and another using a disc with an azimuthal disturbance of mode number 20 (b) and amplitude of $2\%$ of the mean disc radius. In both cases the mean radius of the disc is $\unit{20}{\milli\meter}$ and the impact speed $\unit{1}{\meter\per\second}$. The walls of the cavity in (b) acquire a structure that resembles the skin of a pineapple. This is explained by the oscillations triggered by the azimuthal disturbance.}
\label{fig:axivsnonaxi}
\end{figure}

We study the cavity produced when a round disc with an azimuthal disturbance of its edge is driven downwards through the free surface of a water volume (see figures \ref{fig:axivsnonaxi} and \ref{fig:expsetup}). This is to be contrasted with the detachment of a bubble from a nozzle with a similar disturbance, which is initially determined by the competition of buoyancy forces with surface tension. The latter effectively smooths out large amplitude and high mode number (short-wavelength) perturbations, making nozzle experiments appropriate only for working with small amplitude, long wavelength disturbances. In addition, if the bubble is grown quasi-statically, viscosity can also play a role in smoothing perturbations due to the small Reynolds number of the water flow induced by the injected air. Therefore, this experiment does not allow much variation of the perturbations' mode number and amplitude.

On the other hand, impact-created cavities are characterized by high Weber and Reynolds numbers from the beginning (provided that the collision speed is high enough). Then, viscosity and surface tension play a marginal role in the formation of the cavity and are given no opportunity to erase features created by large-amplitude or high-mode perturbations. There is no initial surface-tension driven stage in the implosion; instead, the expansion of the cavity is opposed by hydrostatic pressure, which eventually starts the collapse, and is then quickly taken over by inertia. On top of this, since the cavities are created on a free surface, there is unobstructed optical access from the top, making it possible to track the shape of the horizontal section of the collapsing cavity.  Disc impact experiments are consequently ideal for experimenting with the influence of geometry in cavity collapse.

The effects of breaking the axial symmetry are clear (figure \ref{fig:axivsnonaxi}). We present experimental results of cavities with disturbances of mode numbers $2$ to $20$ and amplitudes ranging from $1\%$ to $25\%$ of the mean disc radius. The experimental setup is described in \S \ref{sec:exp}. We then explain the axisymmetric collapse model and the theory for the evolution of a perturbation in \S \ref{sec:theory}. Our experimental observations are shown and discussed in \S \ref{sec:expobs}. The collapse of small-amplitude cavities viewed from the top and the side, and the comparison with theory are presented in \S  \ref{subsec:smallamp}.  Finally, we discuss the collapse of high-amplitude shapes in \S \ref{sec:largeamp} and draw general conclusions in \S \ref{sec:conc}.

\section{Experimental setup and procedure}
\label{sec:exp}

\subsection{Setup}
\begin{figure}
\centering
\includegraphics[width=0.8\textwidth]{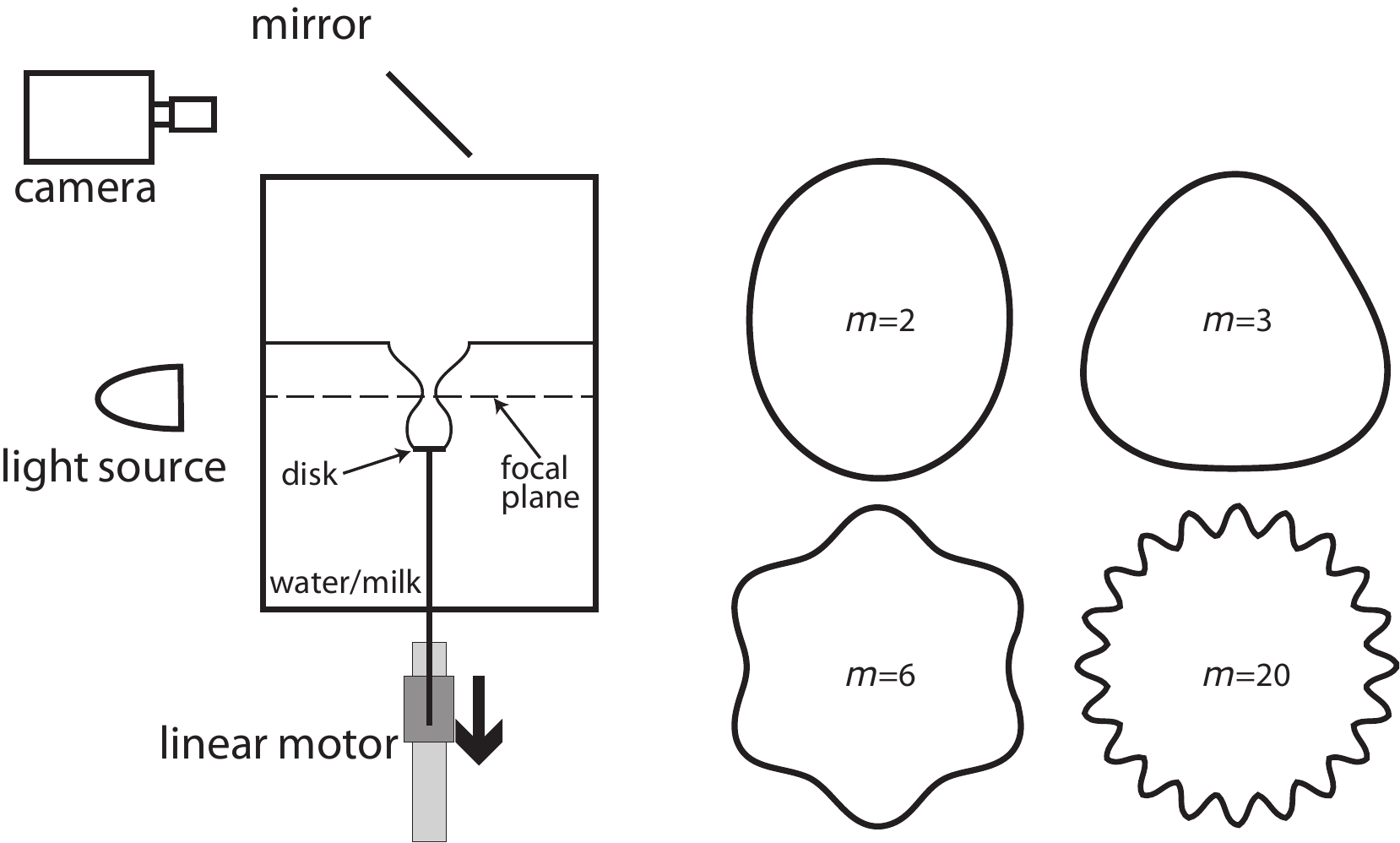}
\caption{Experimental setup and examples of impacting discs. This camera-mirror configuration is used for top views; for side videos we simply move the camera down to the level where the light is. The shown discs have a disturbance amplitude of $10\%$ of the mean radius ($R_{disc}$).}
\label{fig:expsetup}
\end{figure}

The experimental setup consists of a linear motor that drives a vertical thin steel rod downwards through the bottom of a glass tank containing $50\times 50 \times\unit{50}{\centi\meter^3}$ of water (figure \ref{fig:expsetup}). A disc is attached horizontally to the top end of the rod with a mounting system that ensures that the disc is parallel to the undisturbed water surface and remains so throughout the impact. The motor is capable of a maximal $\unit{300}{\meter\per\square\second}$ acceleration and its position can be controlled with a resolution of $\unit{5}{\micro\meter}$ over its $\unit{1}{\meter}$ long track. Since the objects are not dropped into the water, velocity is a control parameter and not a response of the system. The setup allows for a precise control of the impact velocity on the range $\unit{0-5}{\meter\per\second}$. A more detailed description of the apparatus can be found in \citet{Bergmann_JFM_09}.

The shapes of the impacted discs are described by the function$r=S_{disc}(\theta)$, where
\begin{equation}
 S_{disc}(\theta)=R_{disc}+ a_m \cos(m\theta),
 \end{equation} 
 $R_{disc}$ is the mean radius, $a_m$ the disturbance's amplitude and $m$ its mode number  (figure \ref{fig:expsetup}). Discs are machined from a flat stainless-steel plate with a $\unit{2}{\milli\meter}$ thickness. The edges are sharpened to right angles with the intention of pinning the contact line to the lower edge and thus minimize the influence of wetting effects. Great care was taken to ensure that both the disc and the rod were dry before each run, as a single remaining drop on either is enough to noticeably alter the dynamics, especially towards the final instants before pinch-off.

We recorded videos using a Photron SA1.1 high speed camera at frame rates from $5,400$ to $20,000$ fps with resolutions ranging from $1024 \times 1024$ to $512 \times 512$ pixels. We image both top views of the collapsing cavity (focusing on the pinch-off plane) and side views to observe its structure. In order to avoid uneven optical reflections from the surface during top-view experiments, in some of the runs we diluted $\unit{1}{\gram}$ of milk powder per litre of water  and shone light at the liquid, obtaining evenly scattered lighting. In this case the top surface of the disc was covered with black tape for improved contrast. For side views we used water without milk and diffuse illumination from the back. Top view videos were processed to extract the contour of the cavity in every frame and track the amplitude of disturbances as a function of time (\S \ref{subsec:topview}). Side views were directly compared to three-dimensional parametric plots of the modelling equations (\S \ref{subsec:sideview}).

\subsection{Parameters}
The formation of a cavity after impact and its subsequent collapse are determined by the mean radius $R_{disc}$ of the disc, its shape, the impact velocity $V_0$, and fluid properties as kinematic viscocity $\nu$, density $\rho$, and surface tension $\sigma$. Hence, the dimensionless parameters of this experiment are the Reynolds number, $\mbox{\textit{Re}}=V_0 R_{disc}/\nu$, the Froude number $\mbox{\textit{Fr}}=V_0^2/(R_{disc}g)$, and the Weber number, $\mbox{\textit{We}}=\rho V_0^2 R_{disc}/\sigma$. The surface tension (and thus $\mbox{\textit{We}}$) is the only property that varies significantly following the addition of milk for some of the experiments. Since we focus on analysing the influence of the impactor's geometry, the impact speed and the mean disc radius were kept constant in all experiments reported here ($V_0=\unit{1}{\meter\per\second}$ and $R_{disc}=\unit{0.02}{\meter}$). In this way, the two variable control parameters --the mode number $m$ and the amplitude $a_m$--  are related exclusively to the shape of the disc.  In all of our experiments: $\mbox{Re}\sim 2\cdot 10^4$, $\mbox{Fr}\sim 5$, and $\mbox{We}\sim 300-400$ (with measured surface tension values of [$\unit{72}{\milli\newton\per\meter}$] for water and  [$\unit{47.1}{\milli\newton\per\meter}$] for the milk solution, respectively). These values indicate that the dynamics is dominated by inertia from the start of the experiment, and a scaling analysis reveals that this condition prevails throughout the experiment, making it unnecessary to consider their dynamic values. In the end, the only relevant control parameter during the evolution of the cavity is the Froude number since it is the parameter that determines when the cavity enters the inertial collapse regime \citep{Gekle_PRE_09}. 


\section{Models of cavity collapse}
\label{sec:theory}

\subsection{Axisymmetric radial dynamics}
\label{subsec:axisym_dynamics}

The model for an axisymmetric, impact-created cavity collapse by \citet{Bergmann_JFM_09}  neglects vertical flow and its derivatives, assuming the  collapse at each height $z$ to be entirely decoupled from other heights, following previous works \citep{Longuet_JFM_91, Oguz_JFM_93}. Since the flow is considered to be exclusively in the radial direction, the only relevant term from the continuity equation in cylindrical coordinates is:
\begin{align}
\frac 1r \pd{}{r}ru_r&=0
\end{align}
Integration of this equation, with the boundary condition that at the free surface the velocities of the water and the interface must be the same, \textit{i.e.} $u_r(R)=\dot R$ leads to:
\begin{align}
u_r&=\frac{R\dot R}r
\end{align}
which corresponds to a two-dimensional sink flow of strength $Q(t)=R\dot R$, where $R(t)$ is the radius of the cavity, and potential:
\begin{align}
\Phi=Q(t)\ln(r)
\label{eq:flow_potential}
\end{align}

Conservation of momentum is expressed by Euler's equation of inviscid motion:
\begin{align}
-\frac 1{\rho}\pd{P}{r}&=\pd{u_r}{t}+u_r\pd{u_r}{r}
\label{eq:euler}
\end{align}
which must be integrated from the cavity radius $R$ to $R_{\infty}$. The upper integration limit is the length scale where the radial flow has decayed $(R_{\infty}\gg R(t))$; strictly, it should depend on the Froude number and time, but it is possible to determine an approximate constant average value from the experimental conditions and dimensions \citep{Bergmann_JFM_09}. With $\Delta P$ being the (positive) difference between the hydrostatic pressure at $R_{\infty}$ and the atmospheric pressure at the free surface of the cavity and accounting for the Laplace pressure jump ($\sigma/R$) across the interface at $R(t)$, integration of equation (\ref{eq:euler}) yields:
\begin{align}
\Delta P +\frac{\sigma}{R}&=\rho\left[ \frac 12\dot R^2+ \left(\dot R^2+R\ddot{R}\right)\ln\left(\frac{R}{R_{\infty}}\right)\right],
\label{eq:ray-pless}
\end{align}
where terms of $O (R/R_{\infty})$ and smaller were neglected.

In this way, the collapse at each height $z$ is modeled like a two-dimensional Rayleigh-Plesset bubble collapse and the whole cavity is composed of a series of such collapses with different starting times as suggested by \citet{Lohse_PRL_04} in the context of the void collapse in quicksand. The original model does not include surface tension since an analysis of the dimensionless numbers from the problem reveals that surface tension never plays a major role in such a collapse. However, we have included it since disturbing the shape of the cavity creates regions of highly increased curvature where surface tension might play a role.

During the inertial part of the collapse, the logarithmic term in (\ref{eq:ray-pless}) diverges as $R$ goes to zero and thus the only way that equation can remain valid is by having the pre-factor of the logarithmic term go to zero. Integration from time $t$ until the collapse time $t_{coll}$ yields a power law $R(t)=\alpha(t_{coll}-t)^{1/2}$.

Experimental studies have found that the exponent of the power law is higher than $\tfrac{1}{2}$ (typical values found are $0.54-0.60$) \citep{Burton_PRL_05, Thoroddsen_POF_07, Keim_PRL_06, Gordillo_PRL_05, Bergmann_PRL_06, Bergmann_JFM_09} and theoretical studies have shown that the exponent indeed has a weak dependence on the logarithm of the remaining collapse time, approximating to $\tfrac 12$ only asymptotically at the end \citep{Gordillo_JFM_06,Eggers_PRL_07, Gekle_PRE_09}. Nonetheless, the full theoretical result lies remarkably close to a power law fit over many decades in time. Hence, we model the (dimensionless) mean radius $\widetilde R$ of our disturbed collapsing cavities as:
\begin{align}
\widetilde{R} = \alpha\left(\widetilde{t}_{coll}-\widetilde{t}\right)^\beta ,
\label{eq:power_law}
\end{align} 
where the tilde indicates dimensionless quantities obtained by dividing length scales by $R_{disc}$ and time scales by $R_{disc}/V_0$, i.e., $\widetilde{R}(\widetilde{t})=R(t)/R_{disc}$, $\widetilde{t} = tV_0/R_{disc}$ and $\widetilde{t_{coll}} = t_{coll}V_0/R_{disc}$.

\subsection{Azimuthal disturbance}

Memory-encoding vibrations induced by a small geometric disturbance have been predicted theoretically for any mode number $m$, and observed experimentally for cavities with $m=2$ and $m=3$ disturbances, namely bubbles released underwater from a slot-shaped nozzle, by \citet{Schmidt_Nat_09} and \citet{Keim_PRE_11}.  The theoretical model was derived through a perturbation analysis of an azimuthal distortion to the geometry of a cavity with the behaviour described by equation \ref{eq:ray-pless}. A brief explanation of the model follows for the sake of clarity. The complete derivation can be found in \cite{Schmidt_JFM}.

Modelling the flow in an axisymmetric collapse as inviscid, irrotational, and incompressible (using Euler's equations) implies that there is no dissipation of energy. If so, the sum of kinetic and potential energies of the system will be conserved and can be expressed using the Hamiltonian:
\begin{align}
H(R,P_R)=\frac{P_R^2}{2M(R)}+\Delta P \pi R^2+\sigma 2\pi R.
\label{eq:Hamiltonian}
\end{align}
The first term on the right is the total kinetic energy of the moving fluid, expressed in terms of the  effective mass and its momentum, which are respectively:
\begin{align}
M(R)&=2\rho\pi R^2\ln(\frac{R_{\infty}}R) \label{eq:mass}\\
P_R&=M(R)\nd Rt. \label{eq:mmomentum}
\end{align}
The second term is the potential energy due to the pressure difference between the fluid bulk and the cavity (at ambient pressure), and the third term is the energy cost of creating a free surface with the shape of the void. Applying Hamilton's equations of motion $\dot R =\partial H/ \partial P_R$ and $\dot{P_R}=-\partial H /\partial R$ we recover equation \ref{eq:ray-pless}. We are thus faced with a dynamics with one degree of freedom, $R$, and one constant of motion, namely the total energy. The implications of this are important: the problem is integrable, has a perfect memory, and according to the Kolmogorov-Arnold-Moser theorem if such a system is perturbed the new dynamics should closely follow that of the unperturbed situation; \textit{i.e.}, breaking the axial symmetry of the cavity by introducing a small disturbance of the shape should yield a collapse with the same leading order dynamics and new (approximately) conserved quantities, keeping it nearly integrable.

Although in our experiments the disturbance of the shape of the cavity is characterized by a single mode $m$, the theoretical analysis considers a perturbation composed of a sum of Fourier modes $\cos(m\theta)$. For a conveniently chosen origin of the $\theta$ coordinate, the perturbed shape of the void is
\begin{align}
S(\theta,t)=R(t)+\sum_m a_m(t)\cos(m\theta),
\label{eq:shape_disturbance}
\end{align}
where $R(t)$ is the mean cavity radius, which should follow the dynamics of the axisymmetric case, and $a_m(t)$ is the amplitude of each mode, which must be small when compared with the mean radius ($a_m(t)/R(t)\ll 1$) in order for the small perturbation theory to hold. Analysis of how the flow is modified by this shape disturbance \citep{Schmidt_JFM, Schmidt_Nat_09} gives a linear second order ODE for the dependence of time evolution of the amplitudes $a_m$ on the mean radial dynamics:

\begin{align}
\ddot a_m+\left(\frac{2\dot{R}}{R}\right)\dot a_m+\left(\frac{\ddot{R}}{R}(1-m)+\frac{\sigma m(m^2-1)}{\rho R^3}\right)a_m=0.
\label{eq:amp_evolution_vnst}
\end{align}
Equation (\ref{eq:amp_evolution_vnst}) includes the additional influence of surface tension due to the surface disturbances on the right hand side of equation (\ref{eq:shape_disturbance}). We can find an approximate solution by substituting $R(t) = \alpha(t_{coll} - t)^{1/2}$, neglecting surface tension, and solving the resulting Cauchy-Euler equation. In dimensionless form this gives
\begin{align}
\widetilde{a}_m(t)=\widetilde{a}_m(0) \cos\left(\tfrac{1}{2}\sqrt{m-1} \ln\left(\widetilde{t}_{coll}-\widetilde{t}\right)\right) .
\label{eq:amp_evol_solution}
\end{align}
We see that the amplitude should oscillate with a constant magnitude and a frequency that diverges as $\widetilde{t}$ approaches $\widetilde{t}_{coll}$ and that higher mode numbers will oscillate faster.

\begin{figure}
\centering
  \subfloat[]{\label{fig:m310plain}\includegraphics[width=0.3\textwidth]{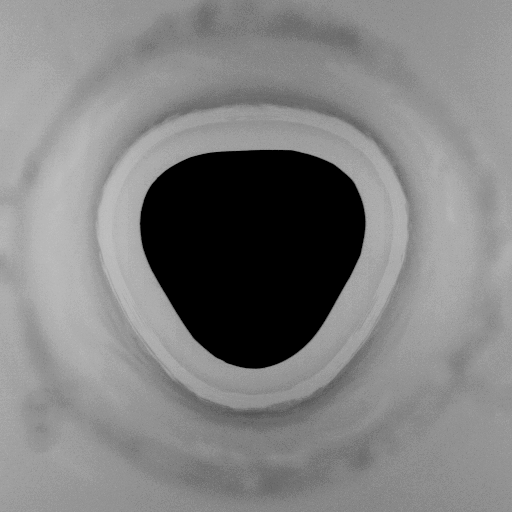}}
  \hspace{.5mm}
  \subfloat[]{\label{fig:m310plain2}\includegraphics[width=0.3\textwidth]{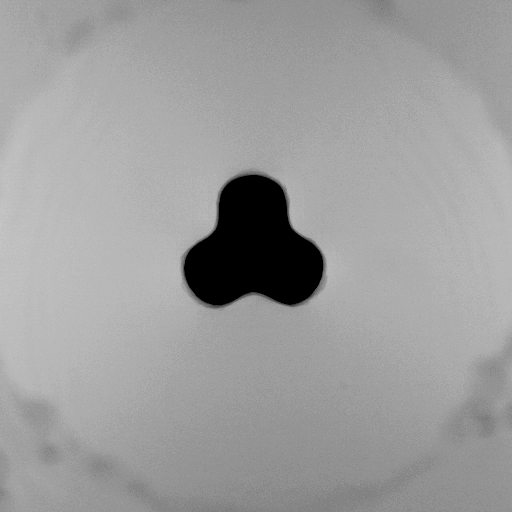}}
  \caption{(movie online) Image of a $m=3$, $R_{disc}=\unit{20}{\milli\meter}$, $a=0.1R_{disc}$ cavity, approx. $\unit{10}{\milli\second}$ after disc impact (a).  Approximately $\unit{90}{\milli\second}$ afterwards, it is clear that the shape has inverted with respect to the original (b).}
  \label{fig:m310collapse}
\end{figure}

Neglecting surface tension, a useful argument to understand the physics of the predicted oscillations is the following:  when the shape of the cavity is disturbed, its curvature is no longer uniform. Therefore, neither is the acceleration of the converging flow associated to its collapse.  As a result of continuity, convergence is stronger in regions with larger curvature, which consequently accelerate more and overtake the regions with smaller curvature (figures \ref{fig:m225collapse} and \ref{fig:m310collapse}), inverting the shape of the cavity. The higher the curvature, the quicker the overtaking becomes; thus with larger mode numbers more oscillation cycles are visible (figure \ref{fig:m162_topview}).

Qualitatively, there is a connection to other instabilities that occur on accelerated fluid interfaces, like the Rayleigh-Taylor (RT) and Richtmeyer-Meshkov (RM) instabilities. Quantitatively this connection is less clear since, instead of being constant (RT) or shock-like (RM), in our case the acceleration is rapidly increasing in magnitude, and even diverges as $\tau\to 0$ as a consequence of continuity.

Equation (\ref{eq:amp_evolution_vnst}), along with that for the axisymmetric radial dynamics (\ref{eq:ray-pless}) are the ingredients for the comparison of the observed cavity shapes obtained from experiments with theory (\S \ref{subsec:smallamp}).


\section{Experimental observations}
\label{sec:expobs}

\subsection{Breaking the axial symmetry: general collapse mechanism}
\label{subsec:break_axial_sym}
\begin{figure}
\centering
\includegraphics[width=0.6\textwidth]{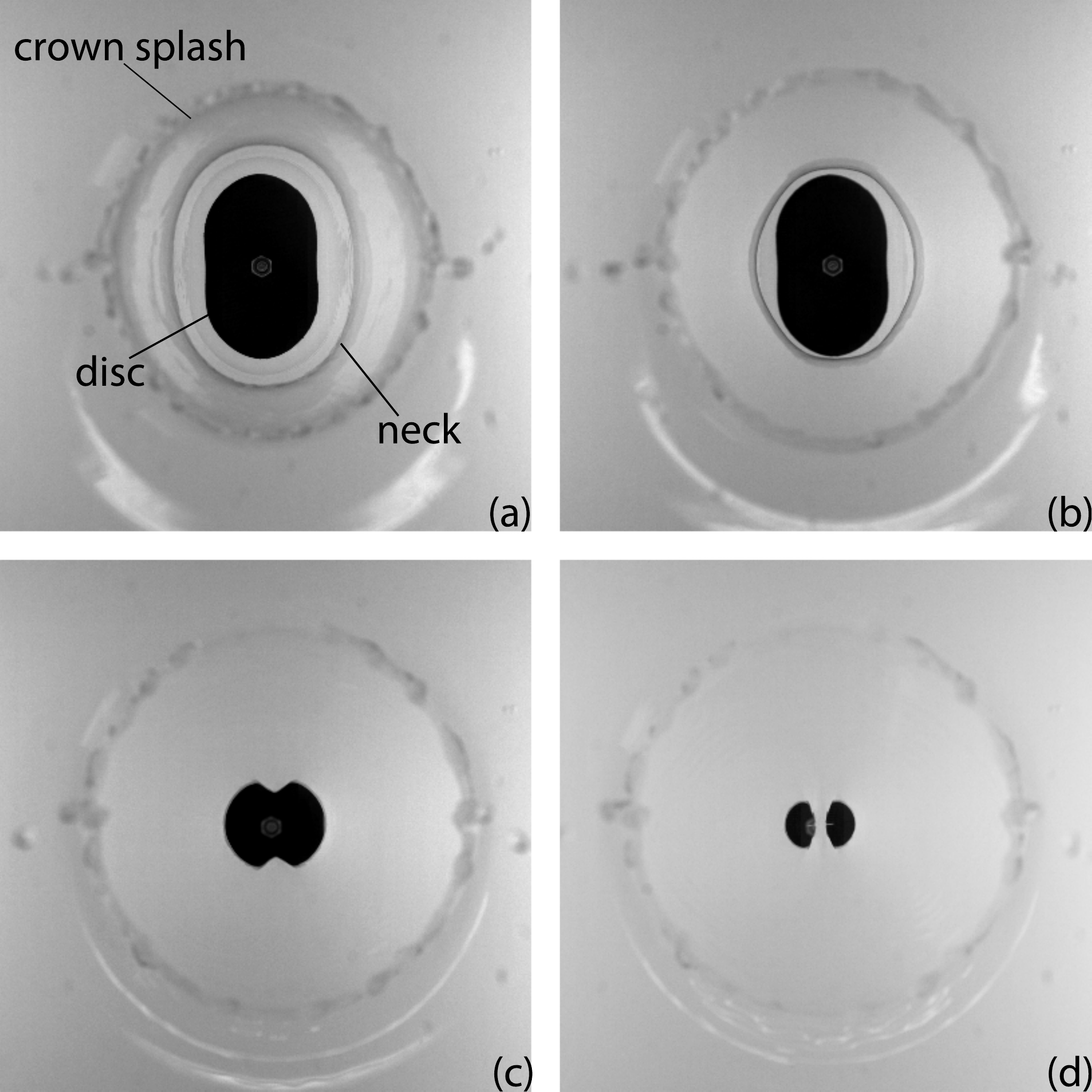}
 \caption{(movie online) Top view of a collapsing cavity ($m=2$, $R_{disc}=\unit{20}{\milli\meter}$ $a_2=0.25R_{disc}$) focused on the pinch-off plane. The cavity (neck) and the disc are initially in phase (a). Since the curvature is higher at the top and bottom of the neck, acceleration along the longer axis is larger. This changes the shape of the cavity as it collapses (b and c). Finally, the two points that were originally farthest apart come into contact first (d).}
  \label{fig:m225collapse}
\end{figure}

Figure \ref{fig:m225collapse} shows the collapse of an elliptical cavity (which can be approximated as a $m=2$ disturbance to a circle) where the longer side of the cavity closes first. Initially, the shape of the cavity  is the same as the impactor that created it (figure  \ref{fig:m225collapse}a). As it closes, the points that were originally farthest apart come towards each other at a higher speed than that of the end points of the minor axis (figure \ref{fig:m225collapse}b). Eventually it becomes clear that the shape of the cavity has inverted its phase with respect to the impactor (figure \ref{fig:m225collapse}c). This shift can be considered as an amplitude inversion of the original shape, described as $S(\theta,t)=R(t)+a_m\cos(m\theta)$ where $R(t)$ is the mean radius and the amplitude, $a$, is originally positive and then becomes negative. 

For small amplitudes (1 or 2\% of the mean radius), oscillations remain linear and with a nearly constant amplitude until very close to the pinch-off moment. However, the fact that we have a disturbance with constant amplitude in a shrinking geometry implies that the disturbance is actually growing with respect to the mean radius and hence it is bound to become of the same order of magnitude at some point. When this happens, the system starts developing higher harmonics and the linear oscillation model can no longer describe the events. At that moment, we say that the collapse evolves into a non-linear behaviour. This was observed for all mode numbers from 2 to 20. In \S \ref{subsec:smallamp} we will first discuss the linear regime, and afterwards turn to the non-linear effects in \S \ref{sec:largeamp}.

\subsection{Effect of small-amplitude disturbances}
\label{subsec:smallamp}

\subsubsection{Events on the pinch-off plane}
\label{subsec:topview}

\begin{figure}
\centering
  \subfloat[$\tau=\unit{17}{\milli\second}$]{\label{fig:m162top_1}\includegraphics[width=0.25\textwidth]{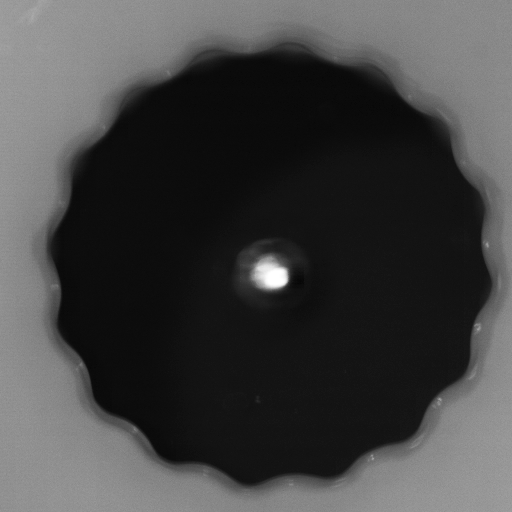}}
  \subfloat[$\tau=\unit{12}{\milli\second}$]{\label{fig:m162top_2}\includegraphics[width=0.25\textwidth]{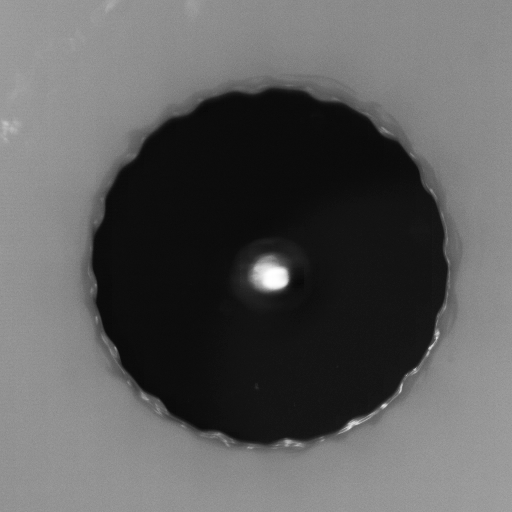}}
  \subfloat[$\tau=\unit{8.4}{\milli\second}$]{\label{fig:m162top_3}\includegraphics[width=0.25\textwidth]{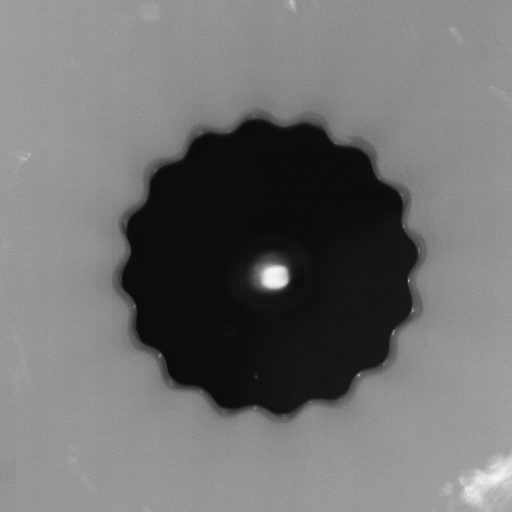}}
  \subfloat[$\tau=\unit{5.5}{\milli\second}$]{\label{fig:m162top_4}\includegraphics[width=0.25\textwidth]{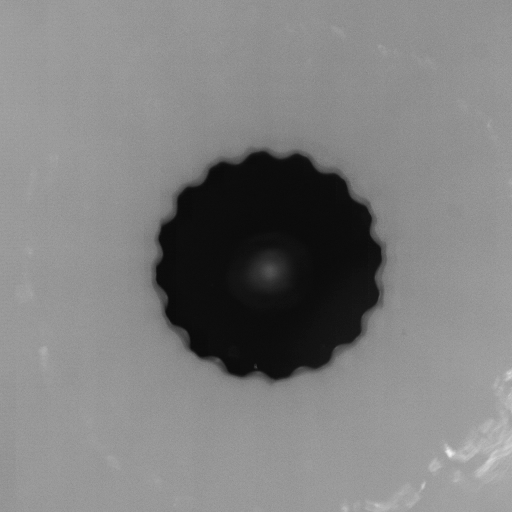}}\\
  \subfloat[$\tau=\unit{2.9}{\milli\second}$]{\label{fig:m162top_5}\includegraphics[width=0.25\textwidth]{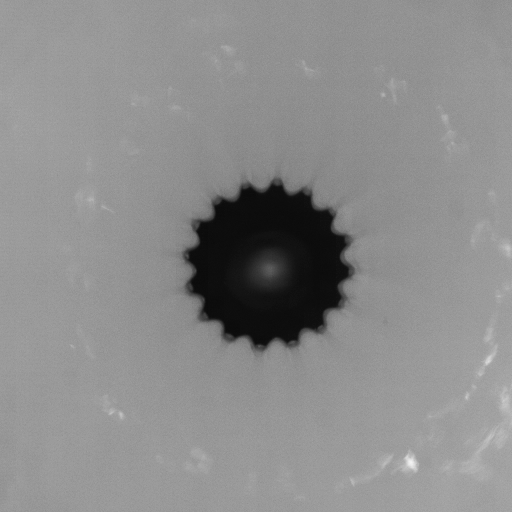}}
  \subfloat[$\tau=\unit{1.4}{\milli\second}$]{\label{fig:m162top_6}\includegraphics[width=0.25\textwidth]{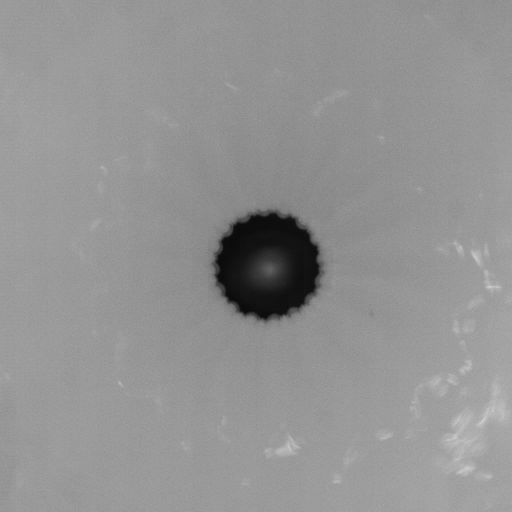}}
  \subfloat[$\tau=\unit{0.45}{\milli\second}$]{\label{fig:m162top_7}\includegraphics[width=0.25\textwidth]{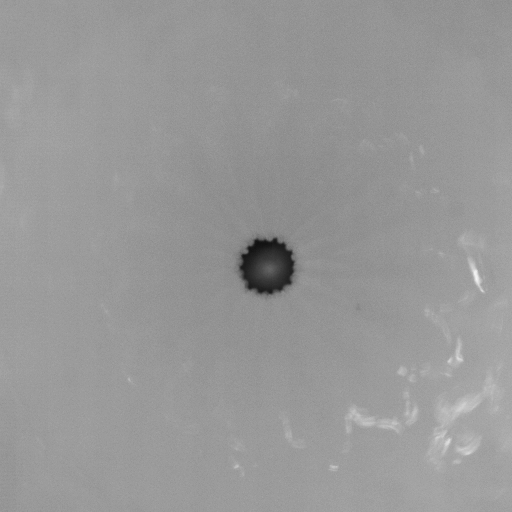}}
  \subfloat[$\tau=\unit{0.15}{\milli\second}$]{\label{fig:m162top_8}\includegraphics[width=0.25\textwidth]{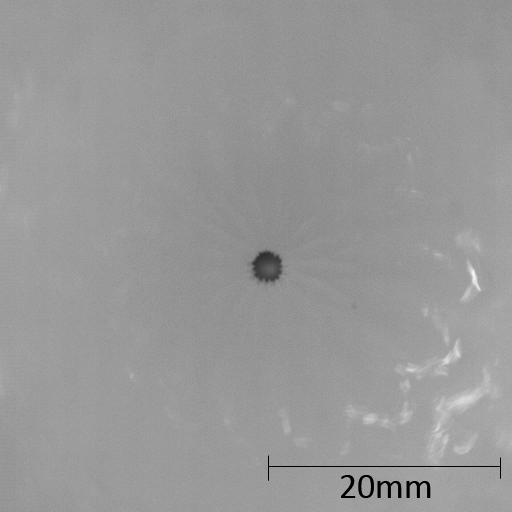}}
  \caption{(movie online) Eight snapshots from a collapse with $m=16$, $R_{disc} = 20$ mm and $a_m = 0.02 R_{disc}$. The 16 peaks and valleys from the original shape can be clearly seen (a). The amplitude decreases on the way to inversion (b) and for an instant has a nearly round shape. The amplitude increases again (c) but it has now inverted with respect to (a). The process carries on (d,e) until we can still see a disturbance but cannot make out the details clearly at our experimental resolution (f,g). Finally, the void pinches off (h).}
  \label{fig:m162_topview}
\end{figure}

Small disturbances do not decay during the collapse of the cavities formed in our experiments. In figure \ref{fig:m162_topview} we show as an example eight snapshots of the evolution of an $m=16$ cavity. Within the bounds set by the used frame-rate and resolution ($20000$ fps and a scale of approximately $86\ \mu\mbox{m/pixel}$), the cavities retain memory of the shape that created them throughout the whole collapse process. We tracked the edge of the cavity $S(\theta,t)$ at every frame from the top-view videos, and found the mean radius $R(t)$ and disturbance amplitude $a_m(t)$ at each time by fitting the curve:
\begin{equation}
S(\theta,t)=R(t)+ a_m(t)\cos(m\theta + \phi_m(t)),
\label{eq:shape_disturbance_fit}
\end{equation}
which is eq. (\ref{eq:shape_disturbance}) for a single mode perturbation. Since $\phi_m(t)$ was found to be nearly constant in time, we chose the reference angle $\theta=0$ to be any line of symmetry in order to match with the model's description of the free surface.

We determined the proportionality constant $\alpha$ and the power-law exponent $\beta$ by fitting equation (\ref{eq:power_law}) to the experimental data of the mean radius. The values found for $\alpha$ were around $1.4$ for all realizations. For a constant sink-flow of dimensionless strength $\widetilde{Q}=\widetilde{R}\dot{\widetilde{R}}$, with $\widetilde R$ given by equation (\ref{eq:power_law}) it follows that $\widetilde Q=\beta\alpha^2\widetilde{\tau}^{2\beta -1}$, where $\widetilde{\tau} = \widetilde{t}_{coll}-\widetilde{t}$. Using $\beta=\frac 12$ (from the power-law model for the axisymmetric case by \citet{Longuet_JFM_91} and  \citet{Oguz_JFM_93}) we find that  $\widetilde Q=\frac{1}{2}\alpha^2$. With $\alpha \approx 1.4$ as found in our experiments $\widetilde Q_{exp}\approx1$. The appropriate dimensional scaling of the flow in this work is $Q_{theo}\sim R_{disc}V_{0}=\unit{0.02}{(\meter^2\per\second)}$, which upon nondimensionalization with the scales $R_{disc}$ and $V_{0}$ becomes $\widetilde Q_{theo}=1$, thereby confirming that the mean flow behaviour in our experiments is similar to the axisymmetric case. Values for $\beta$ were found to lie in the range $0.57-0.60$, which is also consistent with previous works (see \S \ref{subsec:axisym_dynamics}). 

\begin{figure}
\centering
  \subfloat{\includegraphics[width=0.48\textwidth]{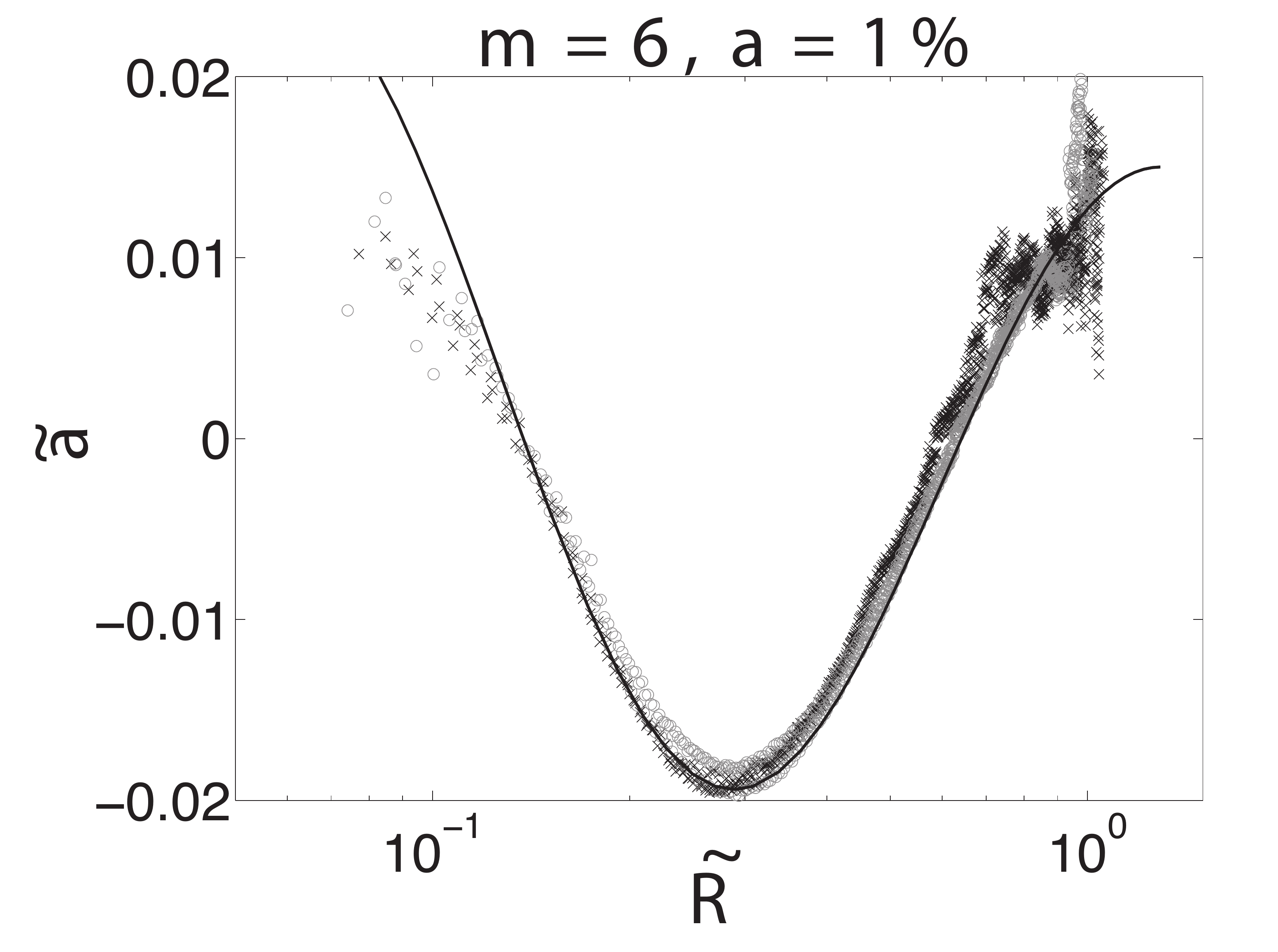}}
  \subfloat{\includegraphics[width=0.48\textwidth]{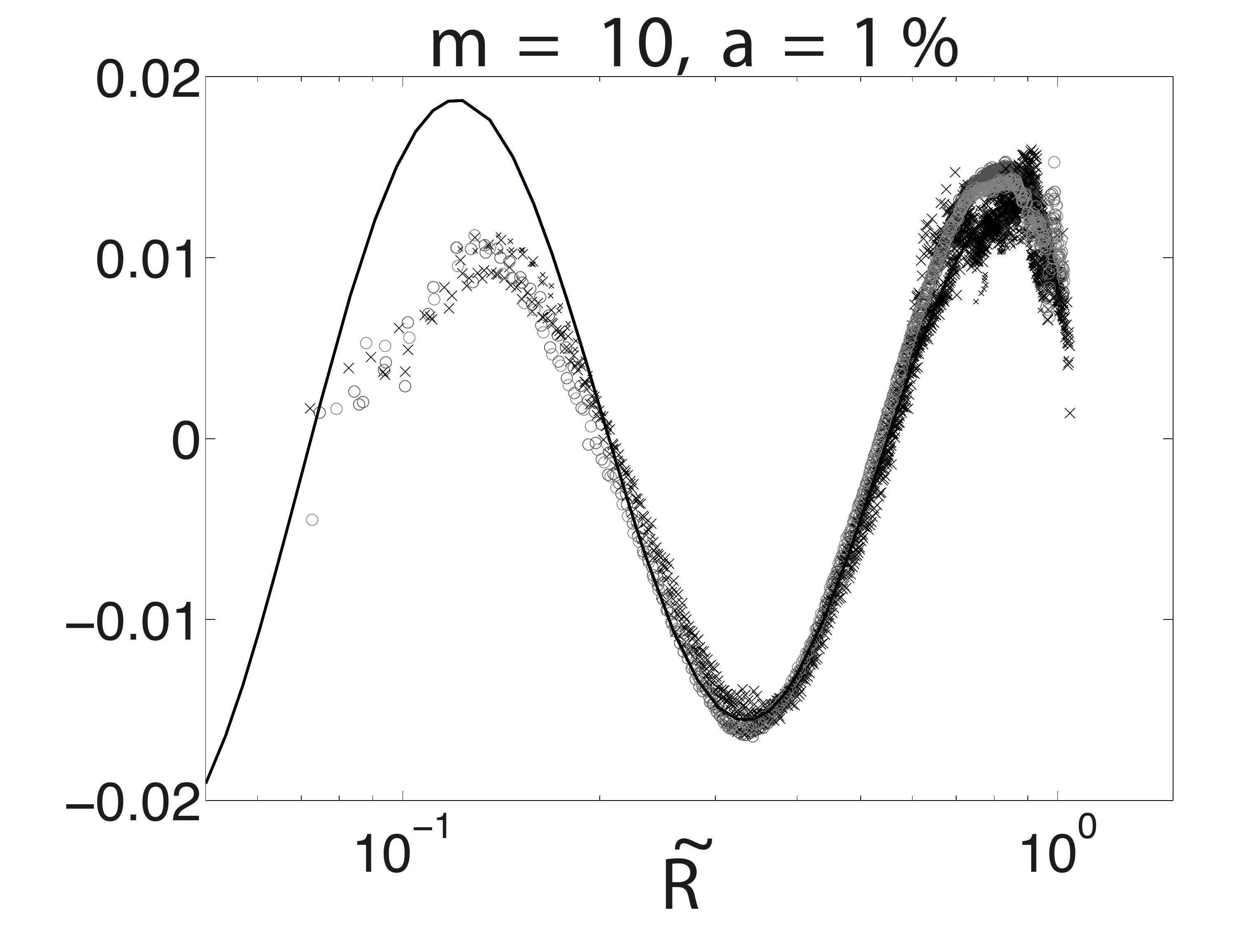}}\\
  \subfloat{\includegraphics[width=0.48\textwidth]{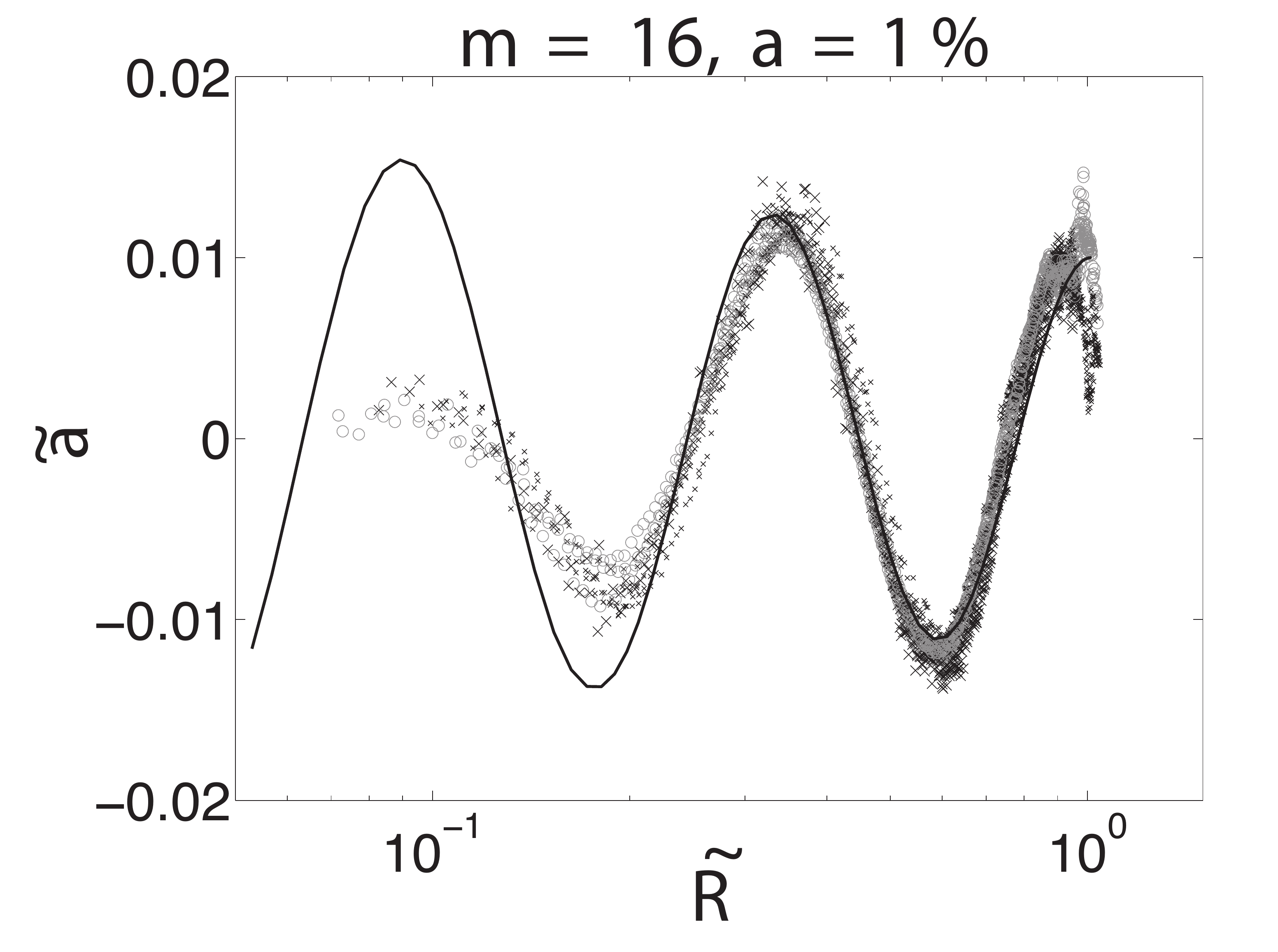}}    
  \subfloat{\includegraphics[width=0.48\textwidth]{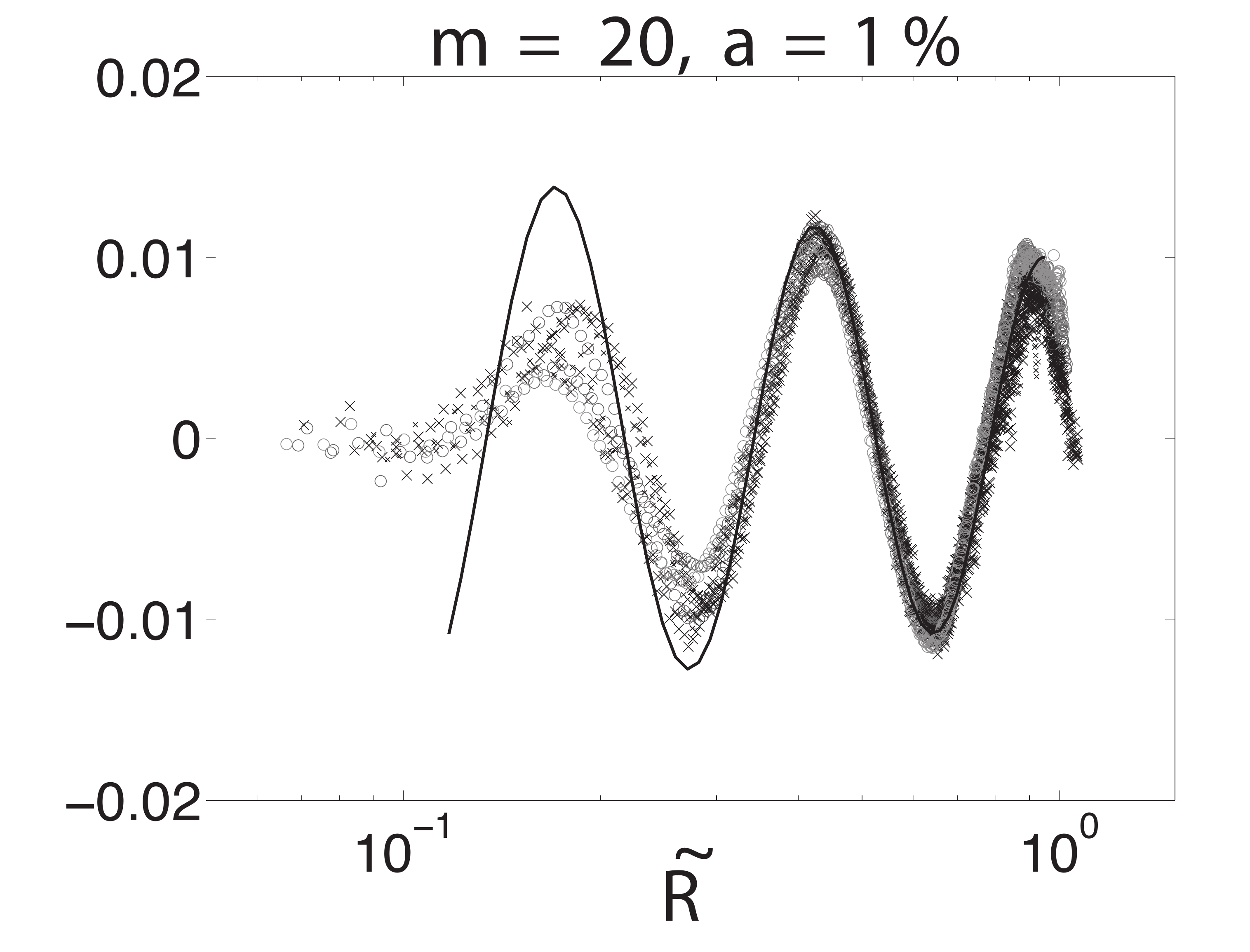}}
  \caption{Evolution of the dimensionless oscillation amplitude as the radius collapses (time goes from right to left). Results for several different mode numbers with an initial amplitude of $1\%$ of the mean radius. Amplitude is expressed as $\widetilde a=a(t)/R_{disc}$. Circles: measurements with milk, crosses: plain water, solid line: theoretical prediction from equation \ref{eq:amp_evolution_vnst} using the measured surface tension for the milk solution and for a given (dimensionless) mean radius $\widetilde{R}=\alpha(\widetilde{t}_{coll}-\widetilde{t})^\beta$ with $\alpha$ and $\beta$ obtained from experiment.}
\label{fig:top_view_graphs}
\end{figure}

\begin{figure}
\centering
  \subfloat{\includegraphics[width=0.48\textwidth]{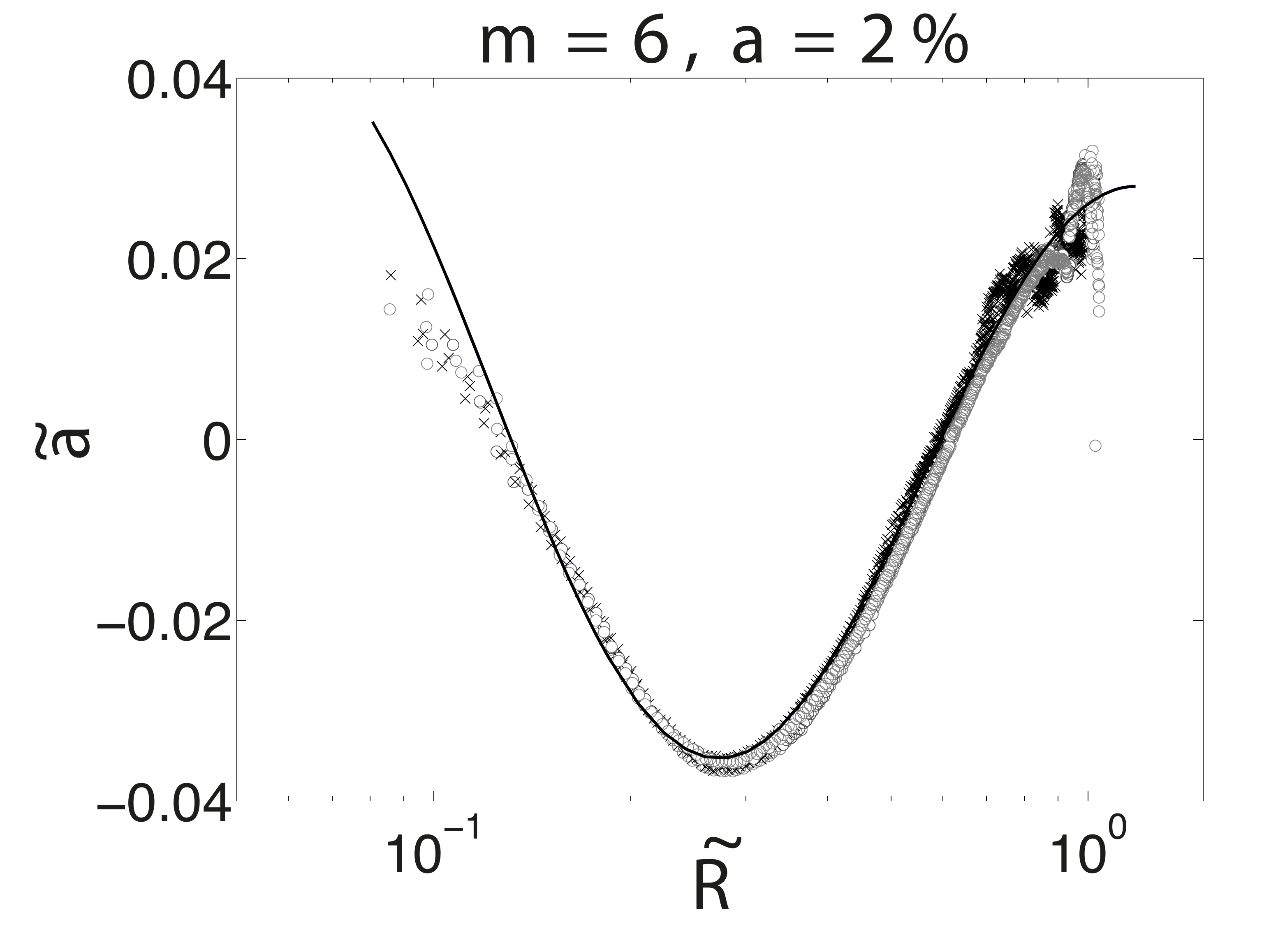}}
  \subfloat{\includegraphics[width=0.48\textwidth]{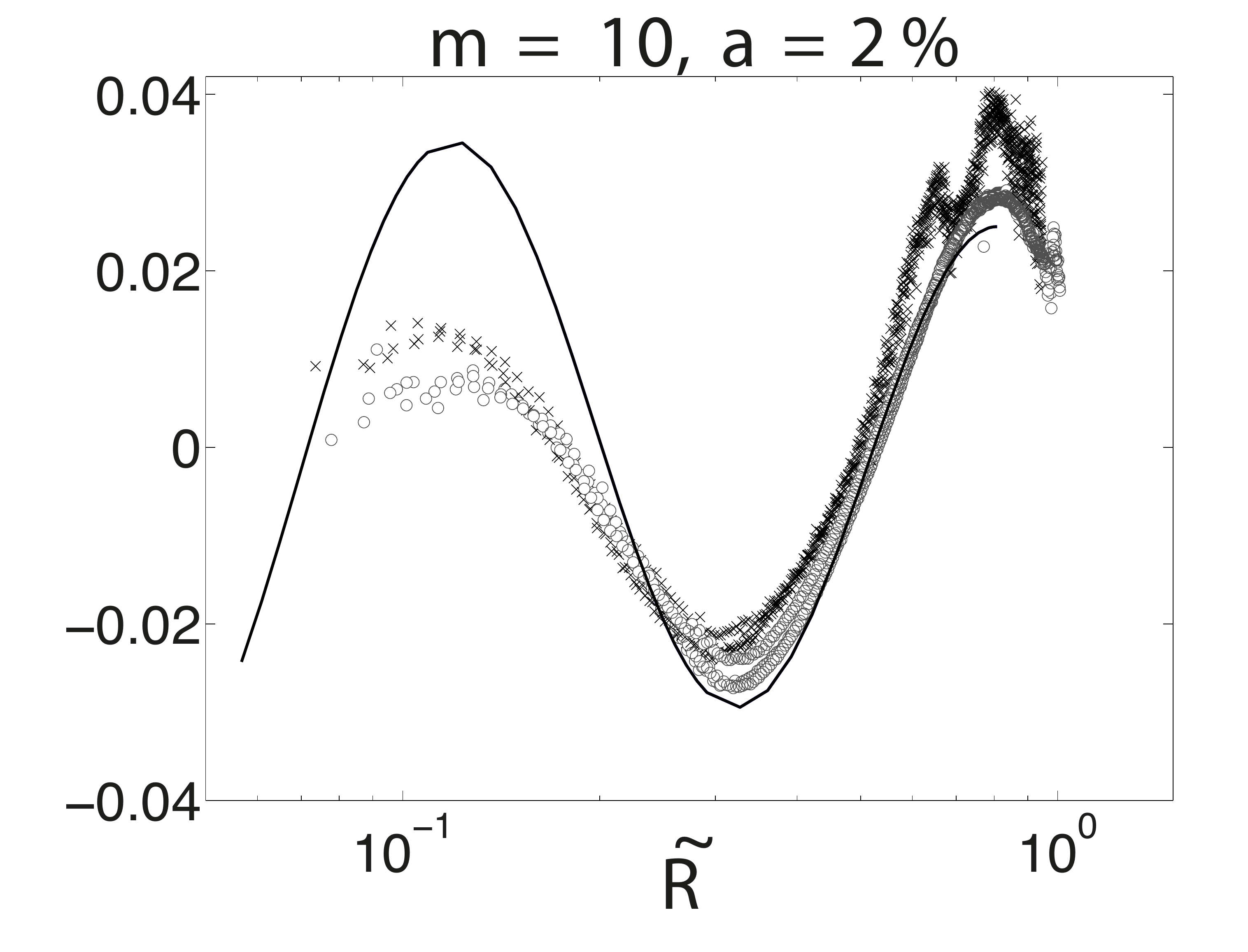}}\\
  \subfloat{\includegraphics[width=0.48\textwidth]{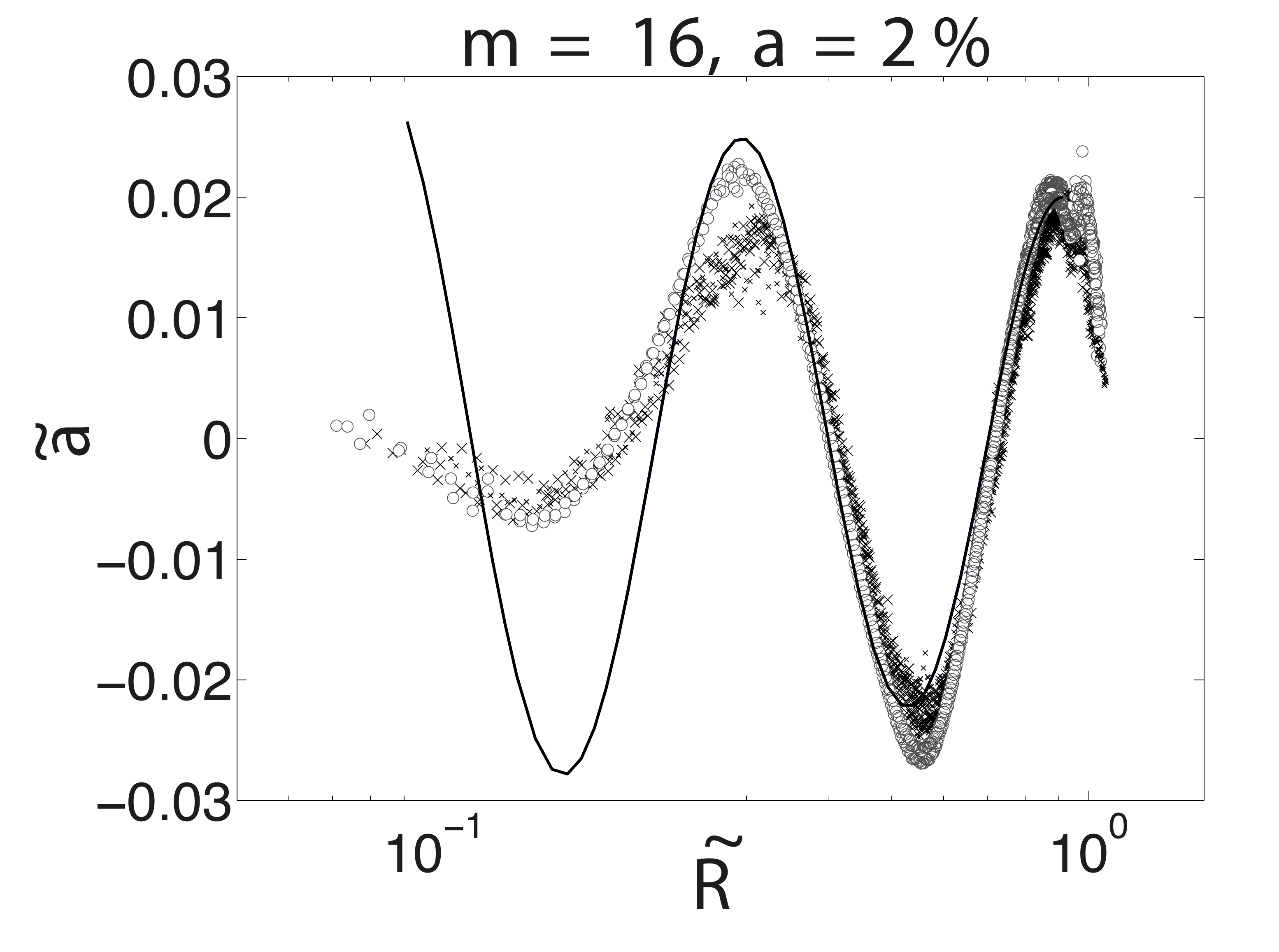}}    
  \subfloat{\includegraphics[width=0.48\textwidth]{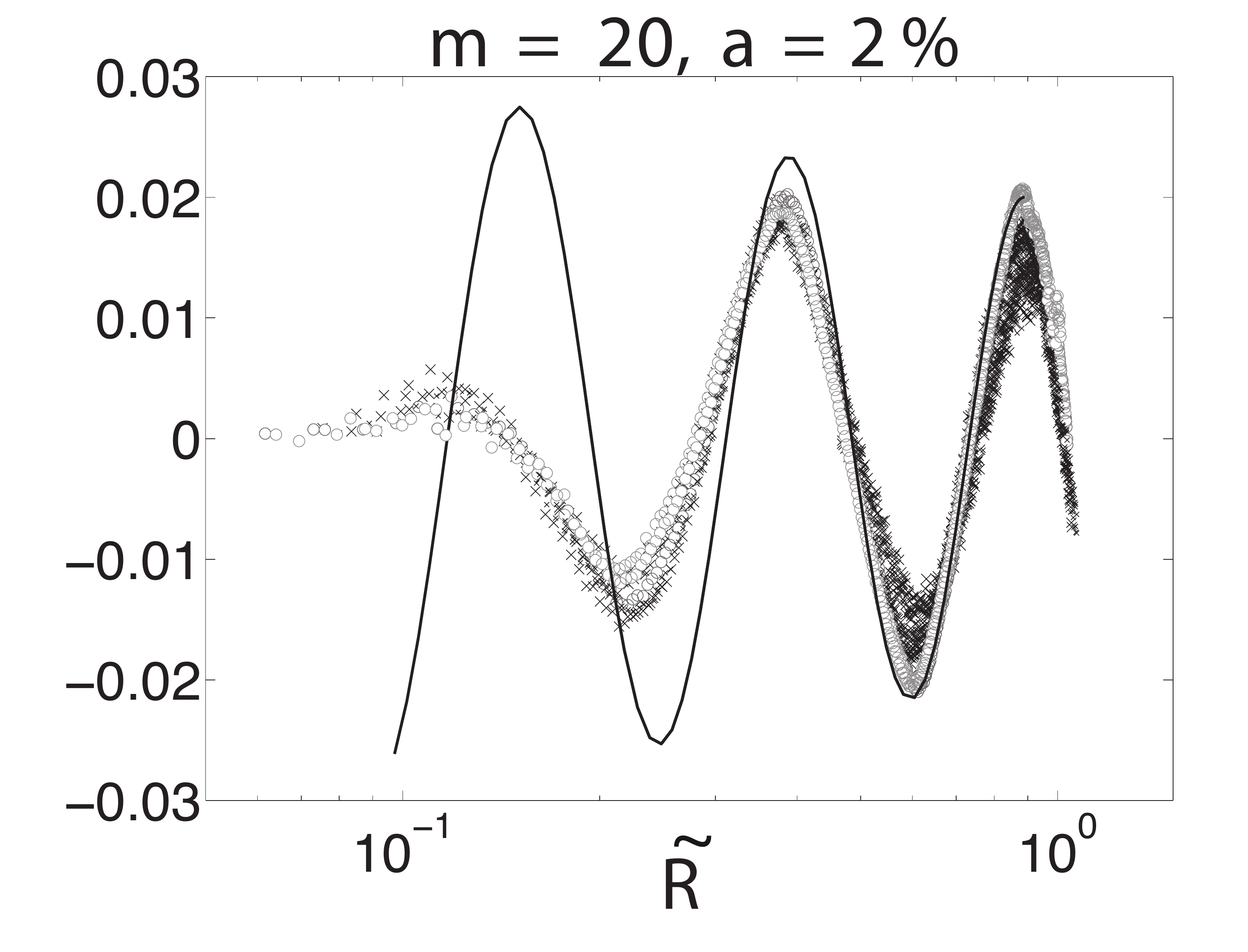}}
  \caption{Evolution of the dimensionless oscillation amplitude as the radius collapses. Results for several different mode numbers with a slightly higher initial amplitude of $2\%$ of the mean radius. Circles: measurements with milk, crosses: plain water, solid line: theoretical prediction.}
  \label{fig:top_view_graphs2}
\end{figure}

Next we constructed the theoretical curves for the evolution of the amplitude $a_m$ by introducing the fitted power-law into equation (\ref{eq:amp_evolution_vnst}). The initial condition for $a_m$ was determined from experimental observations and taken at a maximum of the curve; hence, the initial condition for the derivative is $\dot a_m=0$. Figures \ref{fig:top_view_graphs} and  \ref{fig:top_view_graphs2} show the experimental results compared with the theory in semi-logarithmic plots of the mode amplitude versus the mean cavity radius, nondimensionalized using $R_{disc}$ (in these plots time increases from right to left). The amplitude in the theoretical curves neither blows-up nor decays; it stays roughly constant in time. This is confirmed by the experimental data, at least during the first oscillations. The amplitude in experimental data drops at the end since we lose the capability to faithfully track the edge of the cavity towards the final collapse. 

The oscillation amplitude is roughly preserved, but as $R(t)$ collapses, the relative disturbance $a_m(t)/R(t)$ grows (figure \ref{fig:surface_tension}). Since linear oscillations occur only for $a(t)\ll R(t)$, once this condition is not fulfilled non-linear effects overtake the dynamics, adding complexities to the shape of the cavity and increasing the difficulty of tracing its contour.  The oscillation period looks constant, but as the horizontal axis is logarithmic, the frequency is actually increasing exponentially (chirping, cf. equation (\ref{eq:amp_evol_solution})). This is how the apparent contradiction between the universality of the axisymmetric system and the retention of the azimuthal disturbance at the same time, manifests itself in the dynamics. Part of the information (the mode amplitude) from the initial conditions is encoded and preserved --the cavity `remembers' the shape that created it-- but the chirping of the frequency makes it increasingly difficult to backtrack the evolution of the cavity as the collapse approaches, hence scrambling part of the information at the end \citep{Schmidt_Nat_09}.

\subsubsection{The role of surface tension}
\begin{figure}
\centering
\includegraphics[width=0.48\textwidth]{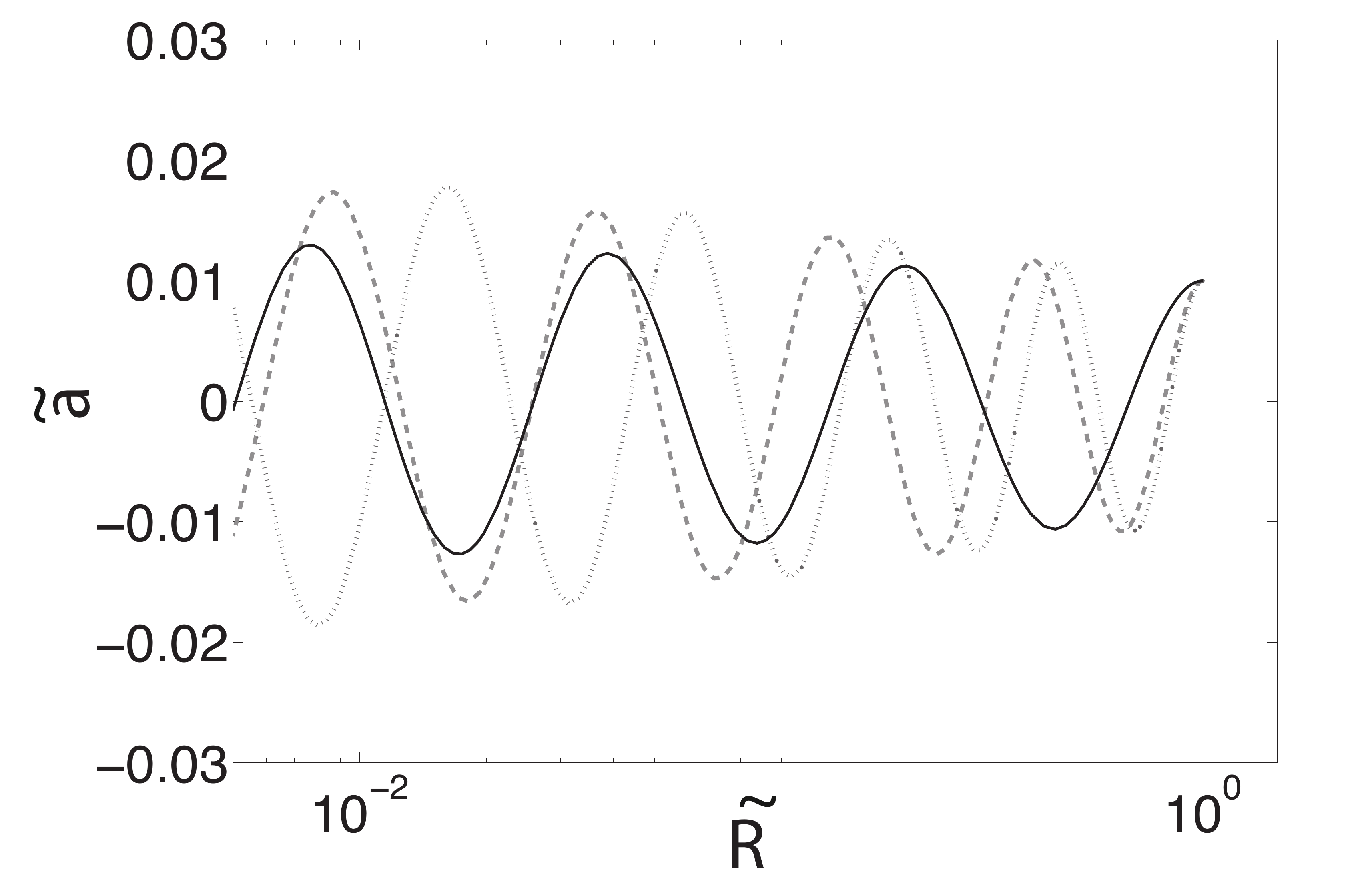}
\includegraphics[width=0.48\textwidth]{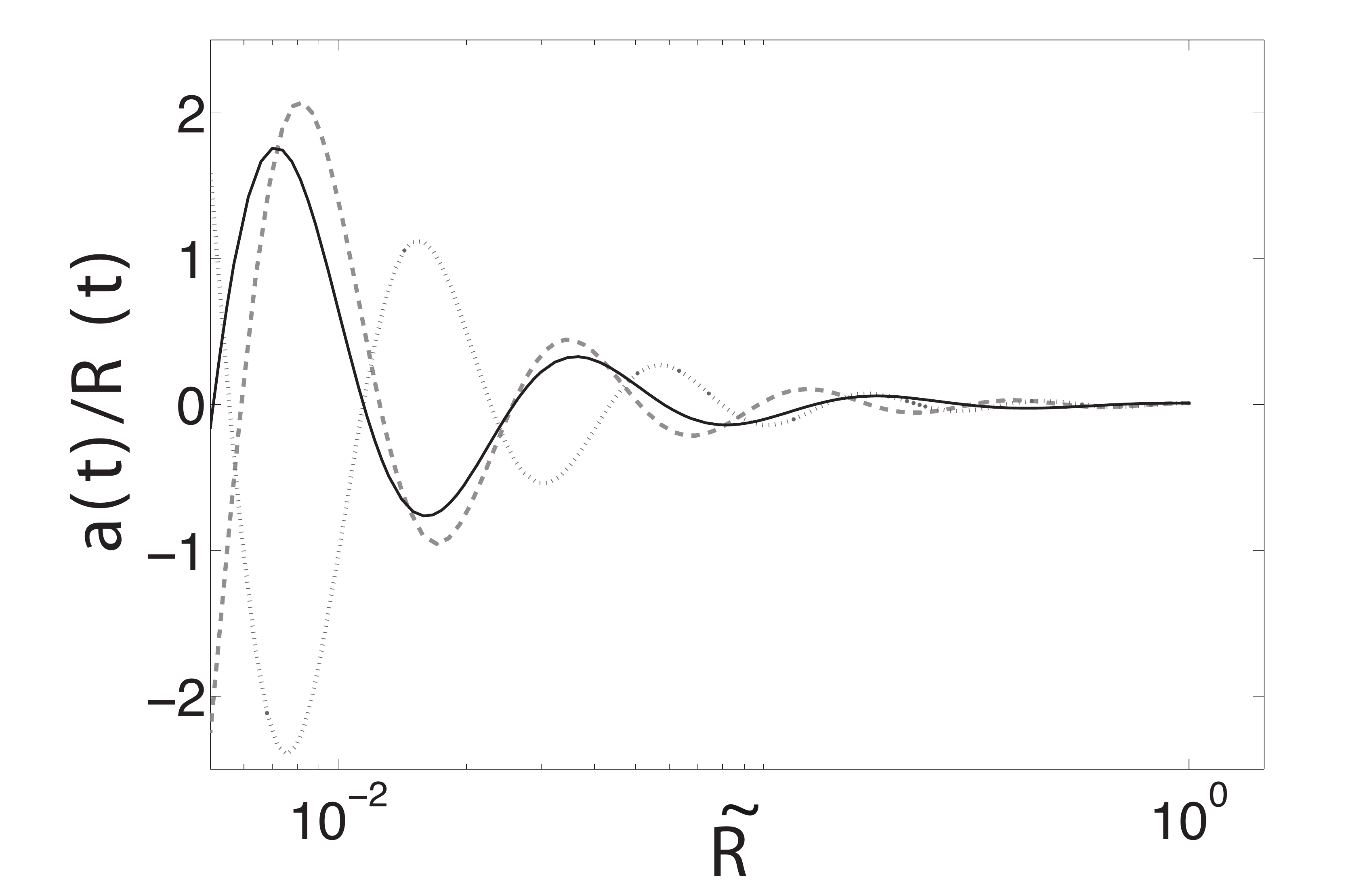}
\caption{Theoretical behaviour of the amplitude, normalized by the disc radius, of a $m=20$, $a_{20}(0)=0.01R_{disc}$ collapse with different surface tension values (left). --- $\sigma=0$,  $--$ $\sigma_m=\unit{47.1}{\milli\newton\per\meter}$ (milk), $\cdots$ $\sigma_w=\unit{72}{\milli\newton\per\meter}$ (water). The same plot is shown on the right, but with the amplitude $a_{20}(t)$ divided by $R(t)$, to illustrate the growth of the relative disturbance.}
\label{fig:surface_tension}
\end{figure}

We shall now, as an intermezzo, comment on the role of surface tension. According to the model, surface tension is expected to change the oscillation frequency and amplitude (figure \ref{fig:surface_tension}). Nonetheless, we see that experimental data acquired using water ($\sigma_w=\unit{72}{\milli\newton\per\meter}$) and data with the milk solution ($\sigma_m=\unit{47.1}{\milli\newton\per\meter}$) lay on top of each other (figures \ref{fig:top_view_graphs} and \ref{fig:top_view_graphs2}). At first this might suggest that surface tension plays no role at all. If this were true, the theoretical curve without surface tension would fit these data; but it does not. Instead, the best fit is obtained with the surface tension of the milk solution. 

The milk powder dissolved in the experimental tank contains surfactant particles which adhere to the free and initially quiescent surface reducing surface tension. When the disc impacts and penetrates, a fresh free surface is rapidly created on the walls of the cavity, to which surfactants take time to adhere. Thus, for the duration of the experiment ($\sim \unit{100}{\milli\second}$) surface tension must be effectively the same as water, or at least considerably higher than the value measured at the static surface. Nonetheless, from fit of the theoretical model we consistently find that the effective surface tension $\sigma_{\textrm{eff}}$ must be lower. We will now show that this can be interpreted as an effect of the axial curvature, neglected by the two-dimensional model.

Considering the axial curvature as being related to the mean cavity radius $R(t)$ (which is the curvature radius in the azimuthal direction) by a (nearly-constant) ratio $\gamma$ such that $R_{ax}(t)=-\gamma R(t)$, we can write the Laplace pressure jump across the interface as $\sigma\kappa=\sigma\left(1/R(t)-1/\gamma R(t)\right)$. The opposing signs in the curvatures are due to the hourglass shape of the cavity, which can be seen in figure \ref{fig:axivsnonaxi}. The right-hand side of the previous expression can be rewritten as $\sigma_{\textrm{eff}}/R(t)$ where $\sigma_{\textrm{eff}}=\sigma\left(1-1/\gamma\right)$, which for any $\gamma>1$ gives a lower effective surface tension, thereby qualitatively explaining our experimental observations. Figure \ref{fig:axivsnonaxi} also makes clear that $R_{ax}>R(t)$ and thus indeed $\gamma>1$. From side-view experimental images, we have observed that during the analysed part of the collapse, $\gamma$ varies between 2 and 4. The relation between the effective surface tension used for the theoretical fit ($\sigma_m$) and the value for water ($\sigma_w$) is $\sigma_m\approx \frac 23 \sigma_w$, corresponding to $\gamma\approx 3$, which is in quantitative accordance with our estimation of the effect of the axial curvature.

\subsubsection{The structure of the cavity}
\label{subsec:sideview}
\begin{figure}
\centering
  \subfloat[]{\label{fig:m204sideimage2}\includegraphics[height=0.23\textheight]{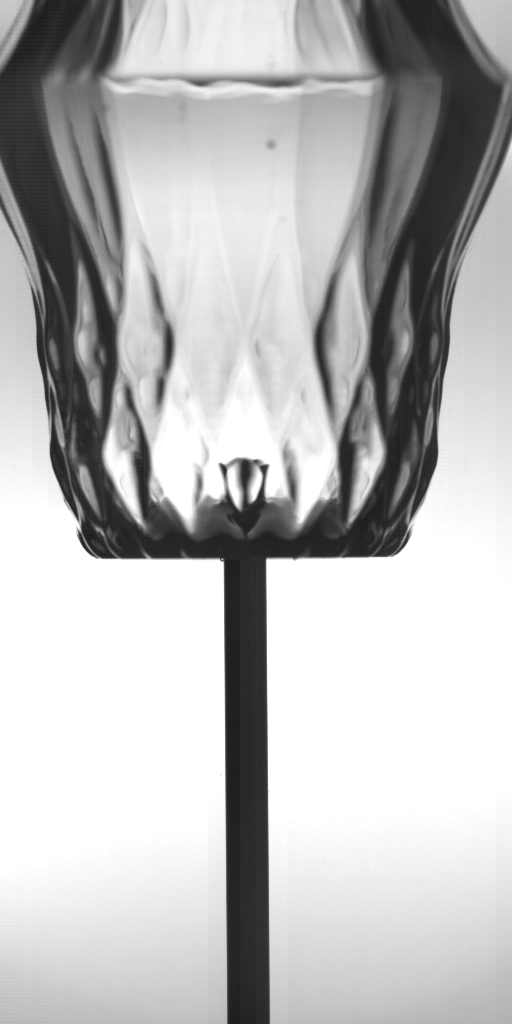}}
  \hspace{.5mm}
  \subfloat[]{\label{fig:m204sideimage3}\includegraphics[height=0.23\textheight]{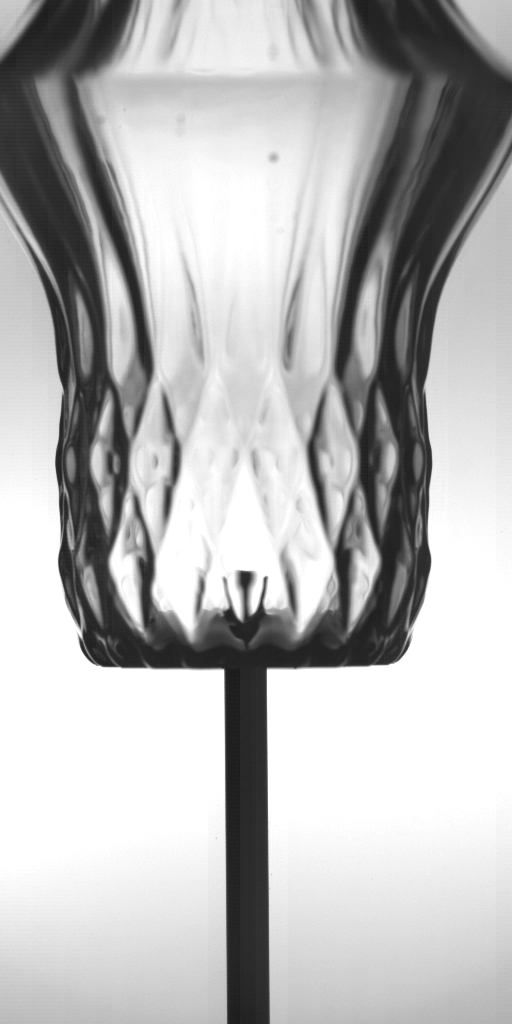}}
  \hspace{.5mm}
  \subfloat[]{\label{fig:m204sideimage4}\includegraphics[height=0.23\textheight]{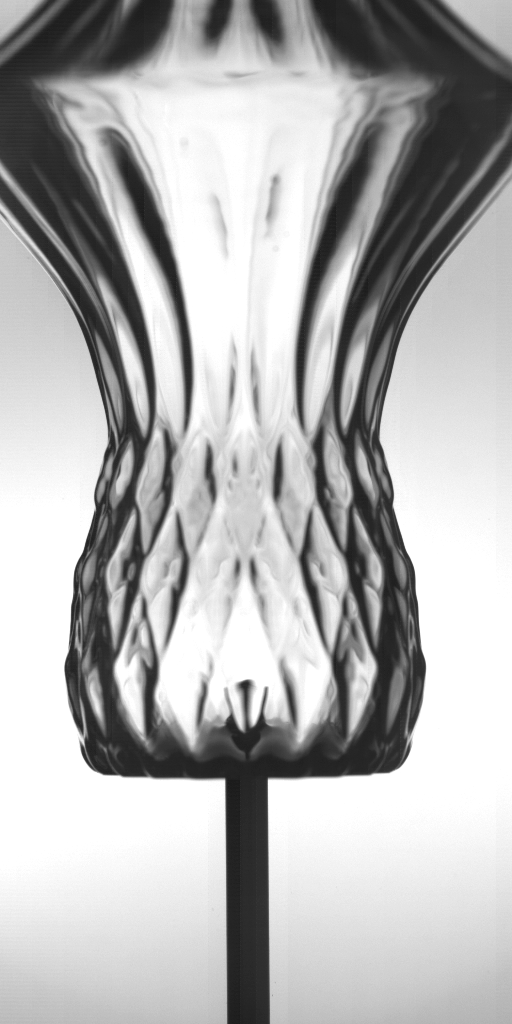}}
  \hspace{.5mm}
  \subfloat[]{\label{fig:m204sideimage5}\includegraphics[height=0.23\textheight]{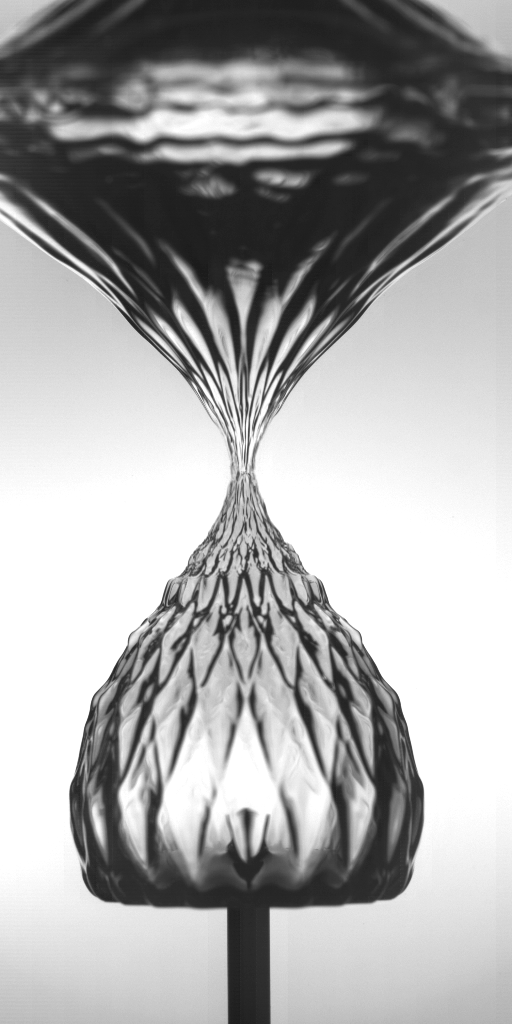}}
  \hspace{.5mm}
  \subfloat[]{\label{fig:m204sideimage6}\includegraphics[height=0.23\textheight]{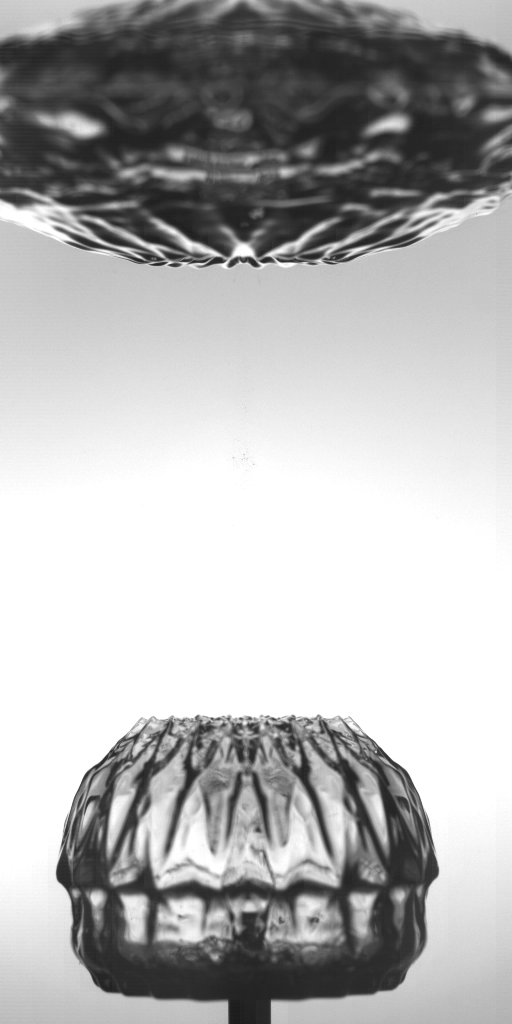}}
  \caption{(movie online) Evolution of the collapse of an $m=20$, $a_{20}=4\%$ cavity as viewed from the side; $R_{disc}=\unit{20}{\milli\meter}$ and $V_{0}=\unit{1}{\meter\per\second}$.}
  \label{fig:m204side}
\end{figure}

We have so far tracked the instantaneous shape of the cavity on a horizontal plane at the pinch-off depth by looking from the top \citep[see][]{Enriquez_POF_10}. Switching to a side-view allows us to see the complete evolution of the cavity at any given time before collapse in a single snapshot. Figure \ref{fig:m204side} shows four images of a cavity before it pinches off and one afterwards. We can see the walls of the cavity developing a structure that resembles the skin of a pineapple \citep[see][]{Enriquez_POF_11}. Furthermore, the structure is not lost after collapse and is still clearly seen in the horizontal cavity ripples that form after pinch-off and were studied in detail by  \cite{Grumstrup_PRL_07}.

\begin{figure}
\centering
\subfloat[]{\label{fig:m202sidemodel}\includegraphics[width=0.3\textwidth]{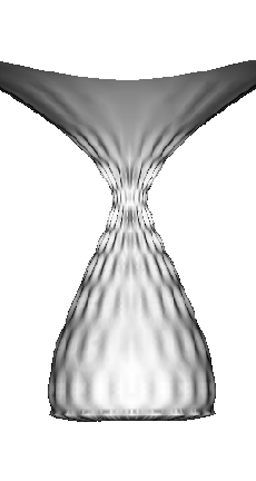}}
  \hspace{5mm}
  \subfloat[]{\label{fig:m202sideimage}\includegraphics[width=0.3\textwidth]{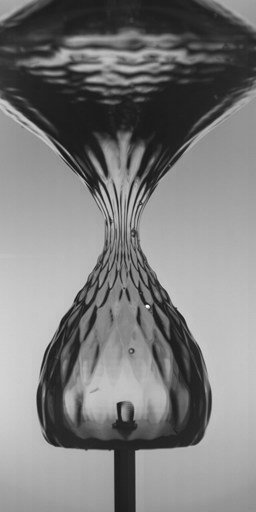}}
  \hspace{5mm}
 \subfloat[]{\label{fig:m202bimodel}\includegraphics[width=0.3\textwidth]{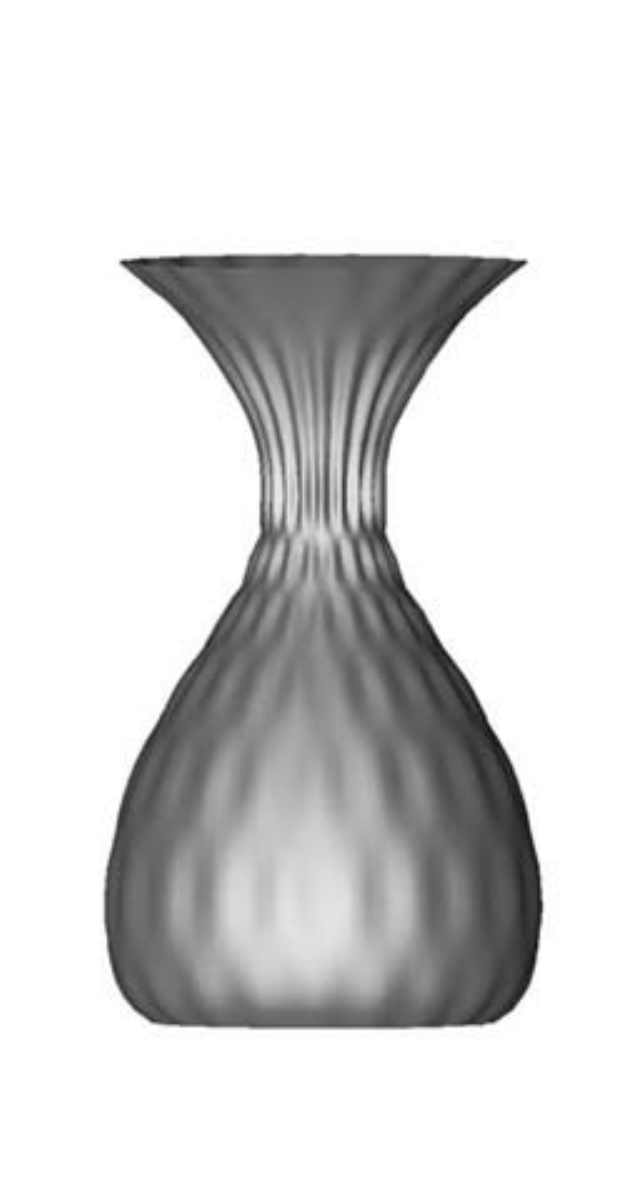}}
  \caption{(movie online) Side view images from experiments and model of $m=20$, $a_{20}=2\%$ collapses. (a) was obtained through the simultaneous solution of the model for two-dimensional axisymmetric collapse and the one for the amplitude of a disturbance; (b) is an experimental image; and (c) was obtained by applying the second model to a cavity profile that resulted from a boundary integral simulation of the axisymmetric case.  This last method is more effective in reproducing the shape of the cavity, as it captures the axial curvature better.}
  \label{fig:m202side}
\end{figure}

We can reconstruct the shape by combining the model for axisymmetric collapse (\ref{eq:ray-pless}) with the equation for the perturbation's amplitude (\ref{eq:amp_evolution_vnst}). First we solve the equation for the mean radial dynamics (\ref{eq:ray-pless}), and afterwards introduce the obtained $R(t)$ and its time derivatives in (\ref{eq:amp_evolution_vnst}).  Since the models are two-dimensional and decoupled in the vertical direction, the three-dimensional shape is built by solving the equations simultaneously at several depths $z$ as done in \citet{Lohse_PRL_04} for the cylindrical void collapse in dry quicksand. Figure \ref{fig:m202side} shows a parametric plot of the solutions just before pinch-off (\ref{fig:m202side}a)  and an experimental image at the same time (\ref{fig:m202side}b). We can improve on this result by using the same axisymmetric boundary integral code that was used in \citep{Bergmann_PRL_06, Gekle_PRL_08, Gekle_PRL_09} to obtain the undisturbed cavity profile $R(z,t)$ which has been found to be in very good agreement with the experimental results \citep{Bergmann_JFM_09} and again use equation (\ref{eq:amp_evolution_vnst}) to superimpose the effect of the disturbance in exactly the same manner as described above. This procedure gives the shape in figure (\ref{fig:m202side}c) which is very similar to the experimental picture, capturing even small details.

\subsection{Effect of large-amplitude disturbances} 
\label{sec:largeamp}

Increasing the amplitude of disturbances gives rise to more complex collapses.  Such cases cannot be described with models derived from the analysis of a small perturbation, since there is a clear non-linear behaviour. Furthermore, the cavities' evolutions are not the same for all geometries. Phase-inversions are still observed and are explained by our continuity argument (\S \ref{subsec:break_axial_sym}) but it is very clear that the cavities no longer close approximately at a single point. Instead, a variety of closure types arise \citep[cf.][]{Turitsyn_PRL_09}; for example: pointy and angular structures, finger-like forms, jets in the radial direction, and sub-cavities. Combinations of two or more of these events might take place. The symmetries of the cavities are always preserved, which means that the original single mode perturbation is dominant. However, other modes are excited and, since symmetry is conserved, the excited mode number $m$ changes to an integer-multiple of its initial value. On top of this, there are situations in which one value of the angular position $\theta$ corresponds to multiple points of the cavity wall, thereby confirming that a simple mode description is no longer applicable.

\subsubsection{Some characteristic features}
\begin{figure}
\centering
\includegraphics[width=0.8\textwidth]{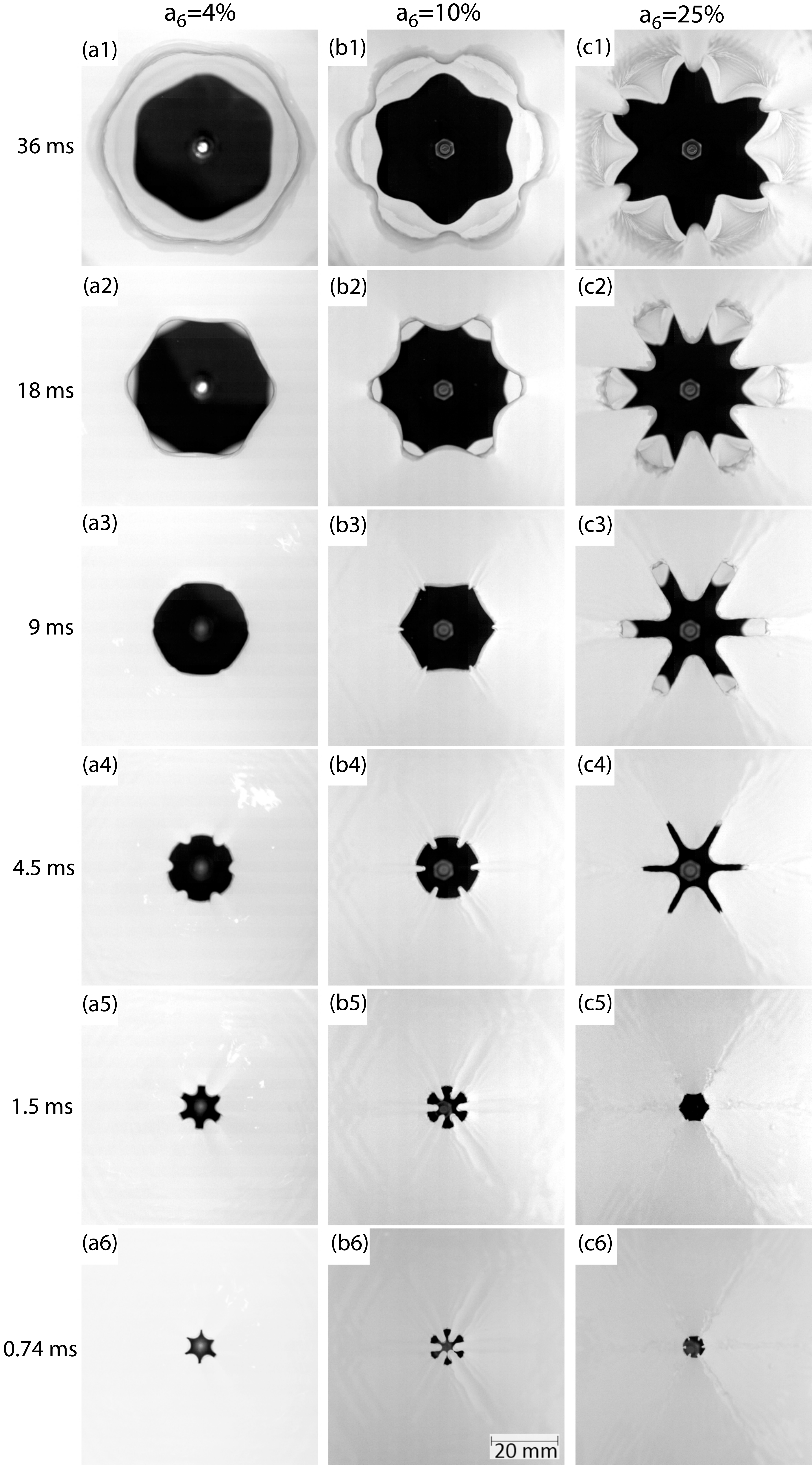}
\caption{(movies online) Comparison of $m=6$ collapses with different initial amplitudes. The oscillation periods of disturbances 4 and 10$\%$ of the mean radius seem a lot more similar to each other than to $a_6=25\%$, where it takes a considerably longer time for the shape to invert. Besides cusps (a6) and jets (b3,c6), we can see finger-like structures (b4, b5) that come in contact to form sub-cavities (b6). When the initial disturbance is very large, such features become very pronounced (series c).}
\label{fig:m6_collapses}
\end{figure}

In figure \ref{fig:m6_collapses} we compare snapshots from $m=6$ cavities with different initial disturbance amplitudes $a_6(0)/R(0)=4\%, 10\%,$ and $25\%$. Each column is a time series of images belonging to one experimental realization. Corresponding snapshots in each series have been taken at equal remaining times to collapse ($\tau$). Whereas the $4\%$ and $10\%$ series are similar, with comparable features occurring at approximately equal times --albeit more pronounced for the larger initial disturbance-- the $25\%$ series differs considerably. A closer look at images \textit{a3} and \textit{b3} ($\tau=\unit{9}{\milli\second}$) in figure \ref{fig:m6_collapses}  reveals pointy, ridge-like jets being formed in singular cusps where the flow converges. For the largest disturbance, this happens only at $\tau=\unit{0.74}{\milli\second}$ (\textit{c6}). In time, such ridge-like formations can evolve  into thicker, finger-like structures (\textit{b4}). 

We found these kind of shapes to be recurrent for most modes and amplitudes but the time at which they happen varies in each case. When they occur too close to pinch-off there is no time for thickening and the collapse takes place with coalescence of the jets in the middle, which tends to happen when the initial amplitude is small ($4\%$ or less). Larger amplitudes seem to be more likely to form bigger sub-cavities. Still, the description cannot be generalized. For example, a $m=6$, $a_6=25\%$ cavity will develop wide liquid structures that seem to crawl towards the center (figure \ref{fig:m6_collapses}, \textit{c1}) and eventually it assumes a shape with thin air arms (\textit{c4}) through which water rushes in to invert the shape; in the end, the collapse (from the top) looks a lot like those with a small initial amplitude. In contrast, a $m=3$, $a_3=25\%$ void will initially form similar thick structures, but they will come in contact with each other, forming large cylindrical sub-cavities (figure \ref{fig:m3subcavities}).

The disturbance amplitude is roughly conserved during non-linear collapse. We have measured the longest and shortest diameters of cavities through several frames and from that estimated the mean radius and amplitude. The maximum amplitudes found have been of the same order as the impacting shape's. In figure \ref{fig:m6_collapses} we can see that the oscillation period changes noticeably when varying the initial amplitude. Look at the row where $\tau=\unit{9}{\milli\second}$ and notice how the two cavities on the left (the smaller amplitude ones) have already started the shape inversion, as made clear by the jets that are being formed, while the rightmost void is still evolving towards the reversal.

\subsubsection{Sub-cavity formation}
\begin{figure}
\centering
\subfloat[$\tau\approx\unit{12.8}{\milli\second}$]{\label{fig:m3sub0}\includegraphics[width=0.24\textwidth]{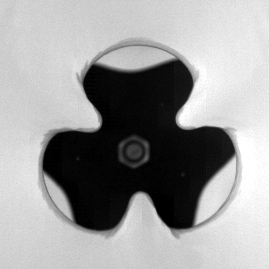}}
\subfloat[$\tau\approx\unit{5}{\milli\second}$]{\label{fig:m3sub1}\includegraphics[width=0.24\textwidth]{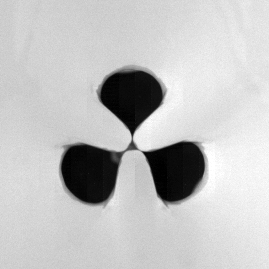}}
\subfloat[$\tau\approx\unit{3.5}{\milli\second}$]{\label{fig:m3sub2}\includegraphics[width=0.24\textwidth]{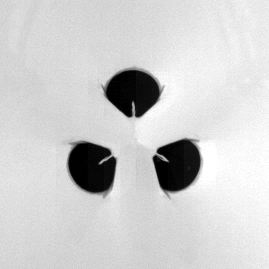}}
\subfloat[$\tau\approx\unit{0.5}{\milli\second}$]{\label{fig:m3sub3}\includegraphics[width=0.24\textwidth]{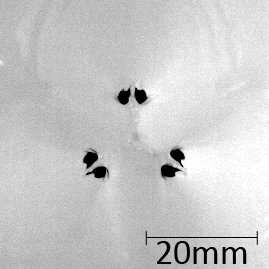}}
\caption[]{(movie online) Example of sub-cavity formation for an $m=3$, $a_3=25\%$ collapse. Three large sub-cavities are formed around a smaller central one which quickly disappears. The inward pointing jets in (a) are due to the growth of the initial $m=3$ disturbance. The jets in (c) are due to the collision (shown in image b) and subsequent convergence of neighbouring regions of fluid which lead to six small cavities before closing (d).}
\label{fig:m3subcavities}
\end{figure}

By looking at the last images before the void closes it is easy to convince oneself that it is very unlikely that the cavities will pinch-off at a single point, thus sub-cavity formation must have happened in every case, although it might be extremely short-lived, not appreciable with our imaging capability, or both. The type of sub-cavity formation depends on whether the last inversion takes place early enough to allow for the thickening of the jet-like structures or not; and we have identified several different cases. For all mode numbers, $m+1$ sub-cavities are formed -a central one surrounded by $m$ satellites- except for $m=2$, where there are just $m$ of these (figure \ref{fig:m225collapse}). Sometimes the central sub-cavity is the smallest and collapses right away while the surrounding ones take longer to disappear. In other cases the exact opposite occurs. An example of the first case is shown in figure \ref{fig:m3subcavities}, where a $m=3$, $a_3=25\%$ void forms $4$ sub-cavities. The center one is gone almost immediately but the other three live long enough to further partition and briefly become $6$ little holes. When we can observe the sub-cavity evolution we always observe a phenomenon like in figure \ref{fig:m3sub1} where small jets are impinged from the contact points into all of the remaining sub-voids.

\subsubsection{Side view}
\begin{figure}
\centering
\subfloat[][Round disc]{\label{fig:roundpinch}\includegraphics[width=0.35\textwidth]{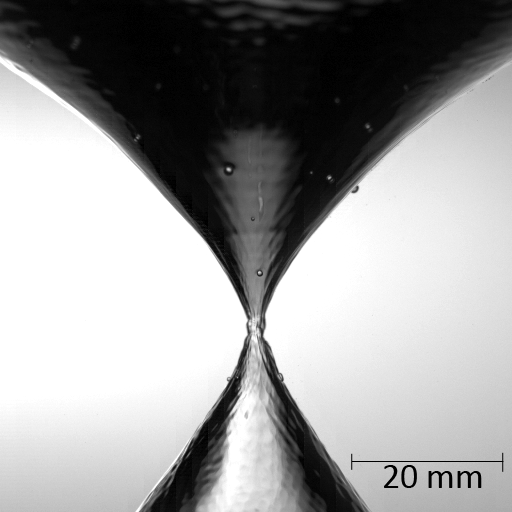}}
\hspace{3mm}
\subfloat[][$m=6,\ a_6=4\%$]{\label{fig:m64pinch}\includegraphics[width=0.35\textwidth]{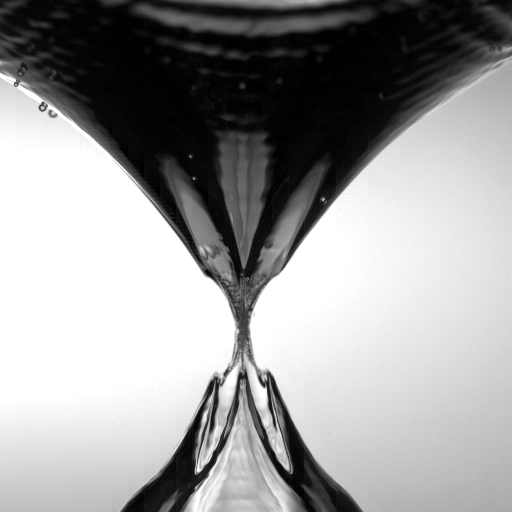}}\\
\subfloat[][$m=6,\ a_6=10\%$]{\label{fig:m610pinch}\includegraphics[width=0.35\textwidth]{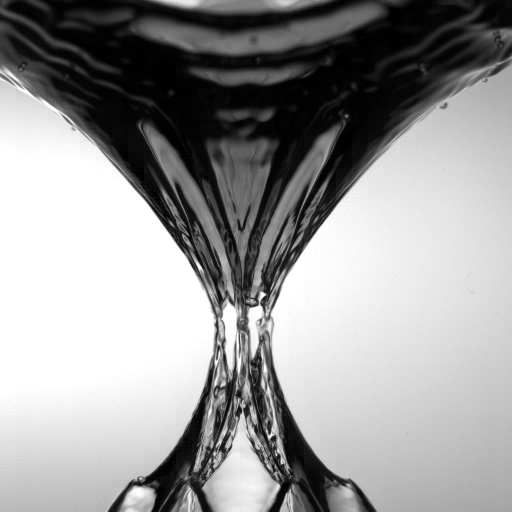}}
\hspace{3mm}
\subfloat[][$m=6,\ a_6=25\%$]{\label{fig:m625pinch}\includegraphics[width=0.35\textwidth]{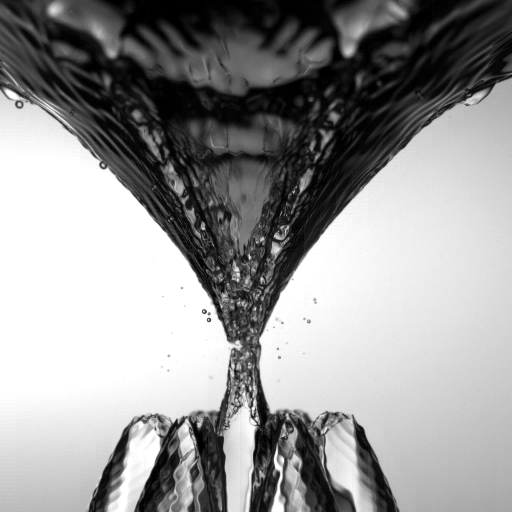}}
\caption[]{(movie online) Effects of axial symmetry breakup on pinch-off. Perturbed cavities no longer collapse at the same point and, as could be expected, the effects become more pronounced with larger amplitudes. Up-down asymmetry is noticeably accentuated, affecting more the void below the deep seal point.}
\label{fig:pinchsidecompare}
\end{figure}

Another difference with small-amplitude collapses is revealed by looking at the pinch-off from the side. Figure \ref{fig:pinchsidecompare} shows the pinch-off moment for a round disc and three $m=6$ discs with amplitudes $4,\ 10,$ and $25\%$. The images in every case are just one frame before the cavity is definitely separated into top and bottom voids. Here the effects of the disconnection not taking place at a single point are manifested in drastic changes of the void's shape at the pinch-off point and accentuation of the top-down asymmetry, which affects the bottom part the most. Figure \ref{fig:m610pinch} shows three air columns with water around them; presumably there are three more columns hidden behind these (see figure \ref{fig:m6_collapses}, \textit{b6}), depicting a situation where seven sub-cavities were formed, the central one collapsed first, and the remaining six close afterwards simultaneously.  In figure \ref{fig:m625pinch} we can see that the pinch-off took place in two stages; the ``arms'' of the cavity closed first, and smaller cavity is left in the center which collapses much like smaller amplitude voids (see third column in figure \ref{fig:m6_collapses}).

\section{Conclusions}
\label{sec:conc}

We have experimentally shown that breaking the axial symmetry of an impact-created cavity leads to oscillations of its walls as it collapses. A small, single-mode disturbance with an amplitude of $1$ or $2\%$ of the mean radius gives rise to linear oscillations of approximately constant amplitude and increasing frequency. The linear behaviour is maintained until the radius shrinks to a size comparable to the disturbance; afterwards, higher oscillation modes evolve and non-linearity sets in. The mean radius evolves in the same way as the axisymmetric, universal case, making this system unique in the sense that it combines universal behaviour in the radial direction with memory of initial conditions in the azimuthal direction. Using two-dimensional models for the mean radius and the disturbance's amplitude we can reproduce the three-dimensional shape of the cavities and understand its structure.

Increasing the amplitude of disturbances induces non-linear behaviour earlier in the collapse. The structures revealed in these cases attest to the beauty underlying collapse phenomena in fluid dynamics. We have observed a variety of pinch-off types arise and vary with mode numbers and amplitudes. Cavities preserve their symmetries in all cases, but the development of higher harmonics in this system is beyond our mathematical modelling so far.

\subsection*{Acknowledgments}
Thanks to Prof. Mark Shattuck for his advice on visualization and image processing, and CONACYT and FOM for providing financial support.


\end{document}